\documentclass[aps,twocolumn]{revtex4}
\usepackage{amsmath}
\usepackage{amssymb}
\usepackage{graphics,graphicx,rotating}

\begin{document}

\title{Appearance of classical Mixmaster Universe \\
from the No-Boundary Quantum State}

\author{Kazuya Fujio}
\email[]{fujio@astr.tohoku.ac.jp}

\author{Toshifumi Futamase}
\email[]{tof@astr.tohoku.ac.jp}

\affiliation{Astronomical Institute,
 Tohoku University, Sendai 980-8578, Japan}

\date{\today}

\begin{abstract}
We investigate the appearance of the classical anisotropic universe from the no-boundary
quantum state according to the prescription proposed by Hartle, Hawking and Hertog. 
Our model is homogeneous, anisotropic, closed universes with a minimally 
coupled scalar field and cosmological constant. 
We found that there are an ensemble of classical Lorentzian histories with anisotropies and experience inflationary expansion 
at late time, and the probability of histories with anisotropies are lower than isotropic histories. 
Thus the no-boundary condition may be able to explain the emergence of our universe. 
If the classical late time histories are extended back, some become singular by the existence of initial 
anisotropies with large accelerations. However we do not find any chaotic behavior of anisotropies near 
the initial singularity.
\end{abstract}


\maketitle

\section{INTRODUCTION}

Quantum cosmology is one of the most ambitious subject in physics. 
Its aim is to treat the universe itself by quantizing spacetime and to explain the creation of universe. 
We have not been able to treat the subject fully satisfactory because we do not have 
a complete theory of quantum gravity. 
However we do have very interesting approaches to the subject using best knowledge of our understanding of general relativity 
and quantum field theory. One is the quantum cosmology based on canonical quantization of general relativity.  
By using Hamiltonian formulation of general relativity and regarding 3-metric and its conjugate as fundamental operators, 
one can obtain Schrodinger like equation on the space of 3-metric(superspace), 
known as {\it Wheeler-DeWitt equation} ${\hat H} \Psi(q^A)=0$, where the $q^A$ are a set of coordinates for minisuperspace and 
$\Psi(q^A)$ is interpreted as wave function of the universe. 
By restricting the spacetime to homogeneous and isotropic(called minisuperspace), Wheeler-DeWitte equation greatly simplifies 
and an interesting idea for the origin of the universe have been proposed\cite{Vilenkin}. 
However, nobody knows how to calculate the probability of specific universe 
in this approach, and there are some problems e.g. the problem of time. 
There has been some progress in the past 10 years or so in this approach based on 
loop quantum gravity(e.g. Ashtekar \cite{Ashtekar}, Bojowald
\cite{Bojowald}).\\
\indent The another approach is based on the path integral quantization and is called no-boundary proposal for 
the wave function of the universe\cite{HH83}. 
In this approach the wave function of the universe is defined by 
summing all Euclidean compact geometry weighted by  
exponential of action $exp(-I)$ without boundary. 
As Hawking said ; {\it the boundary condition of the universe are 
that has no boundary} (Hartle \& Hawking \cite{HH83}, Hawking \cite{Hawking}).
According to this proposal the origin of the quasiclassical realm from the no-boundary proposal for the quantum state of the universe 
has been discussed by Hartle, Hawking and Hertog \cite{HH08} in a class of minisuperspace models with homogeneous, isotropic 
closed spacetime geometries. They argued that a wave function with the semiclassical form 
\begin{equation}
\Psi(q^A)=A(q^A)e^{iS(q^A)/\hbar}
\end{equation}
in some region of minisuperspace predicts an ensemble of classical trajectories provided that $S(q^A)/\hbar$ 
is rapidly varying and $A(q^A)$ is slowly varying. 
It was shown that the complex solutions(metrics) of Euclidean Einstein equations that extremize the Euclidean action  
are distinct from the real Lorentzian classical metrics for which they provide the probabilities, and 
the probabilities for classical histories derived from the no-boundary wave function do define a measure on classical phase space. 

It will be very interesting to see if it is possible to apply their method to more general class of minisuperspace.  
This is the aim of the paper. Namely we investigate the emergence of classical universes with
anisotropy according to the prescription proposed by HHH.
We consider the Bianchi type IX spacetime which is homogeneous but not isotropic,
with a minimally coupled scalar filed moving
in a quadratic potential and cosmological constant.  
Quantum cosmology of Bianchi type IX model has been studied using Wheeler-DeWitte equation.
For example, Battisti et al. \cite{MRG} investigated the situation where the isotropic scale factor is considered 
as intrinsically different variable from anisotropies. Namely, they regarded the scale factor as
time-parameter, and showed that the wave function of the universe is spread over all values of 
anisotropy near the cosmological singularity. However, it is also shown that the wave function asymptotically peaked around 
the isotropic configuration when the universe grows large. 
We will also find that the universe approaches isotropic configuration, but 
possible range of anisotropy near the bounce or singularity is very restricted unlike their result, but 
is consistent with the prediction by loop quantum cosmology\cite{BDH}.  

The organization of the paper is as follows. 
We briefly review mixmaster universe in Sec.~II and 
explain the condition of classicality according to HHH 
in Sec.~III. Then we calculate the wave function of our model and 
applies the prescription of HHH in Sec.~IV. The numerical results are shown 
in Sec.~V. Finally summary and some discussion will be given 
in Sec.~VI.

 
\section{MIXMASTER UNIVERSE}

\noindent In this section we briefly review Bianchi type IX model. 
The geometry of Bianchi type IX model may be 
written as  
\begin{equation}
ds^2 = -N^2dt^2 + e^{2\alpha}(e^{2\beta})_{ij}\sigma^i\sigma^j \\
\end{equation}
where $N$ is the lapse function and $\beta_{ij}$ represent spacetime
anisotropy. The 1-forms $\sigma^i$ are defined by 
\begin{eqnarray}
\sigma^1 &=& \cos \psi d\theta - \cos \psi \sin \theta d\phi\nonumber\\
\sigma^2 &=& \sin \psi d\theta - \cos\psi\sin\theta d\phi\\
\sigma^3 &=& d\psi + \cos\theta d\phi \nonumber
\end{eqnarray}
$N,\alpha,\beta_{ij}$ are functions of time $t$. 
The anisotropy variables $\beta_{\pm}$ are parametrized as follows;
\begin{eqnarray}
\beta_{11}&=&\beta_+ + \sqrt{3}\beta_-\nonumber\\
\beta_{22} &=& \beta_+ - \sqrt{3}\beta_-\\
\beta_{33} &=& -2\beta_+ \nonumber
\end{eqnarray}
\noindent Thus the dynamical variables of our model are $\alpha, \beta_\pm$.  
This model is reduced to closed FRW model when $\beta_{\pm} = 0$. 
The (Lorentzian) action of this model is calculated to be 

\begin{equation}
S = \frac{\pi}{G}\int dt Ne^{3\alpha} \left[\frac{6}{N^2} \left
(-\dot{\alpha}^2 + \dot{\beta_+}^2 + \dot{\beta_-}^2 \right)
+{}^3R_\mathrm{IX} \right] \label{Laction}
\end{equation}

\noindent where we do not include cosmological constant 
at the moment, the three-curvature ${}^3R_\mathrm{IX}$ is 
\begin{eqnarray}
\begin{split}
{}^3R_\mathrm{IX} =& \frac{1}{2} e^{-2\alpha} 
tr(2e^{-2\beta}-e^{4\beta}) \nonumber\\
=& e^{-2\alpha} [ 2e^{-2\beta_+} \cosh{2\sqrt{3}\beta_-}\\ &-
 \frac{1}{2} e^{-8\beta_+} + e^{4\beta_+}(1-\cosh{4\sqrt{3}\beta_-}) ] \\
\equiv& - \frac{1}{2} e^{-2\alpha} V(\beta_+,\beta_-) \nonumber
\end{split}
\end{eqnarray}

\noindent Then the Hamiltonian for the Bianchi type IX universe is
\begin{equation}
H_\mathrm{IX} = \frac{GN}{24\pi} e^{-3\alpha} [ -p_\alpha^2 +
 p_+^2 + p_-^2 + \frac{12\pi^2}{G^2} e^{4\alpha}
 V(\beta_+,\beta_-) ] \label{B9H}
\end{equation}
\noindent The conjugate momenta of $\alpha$ and $\beta_\pm$ are as follows.
\begin{eqnarray}
p_\alpha &=& -\frac{12\pi}{GN}e^{3\alpha}\dot{\alpha}\\
p_\pm    &=& \frac{12\pi}{GN}e^{3\alpha}\dot{\beta_\pm}
\end{eqnarray}

\noindent It is possible to think the Hamiltonian (\ref{B9H}) 
as that of a particle with triangular shaped potential $V(\beta_+,\beta_-)$ whose walls are very steep. 
Therefore the evolution of the closed universe with anisotropy may be expressed by the motion of this point mass 
(e.g. Misner,Thorne \& Wheeler \cite{grav}). Since the motion of the point shows chaotic behavior in general by bouncing 
between potential walls, it is argued that the classical universe is predicted to behave chaotically taking all possible values 
of anisotropies. This model of the universe is called the mixmaster universe .


\section{NO BOUNDARY WAVE FUNCTION OF THE UNIVERSE AND SEMI-CLASSICAL APPROXIMATION}

In this section we review the semi-classical approximation of
the no-boundary wave function of the universe according to HHH. \\
\noindent The no-boundary wave function is defined by sum-over-histories

\begin{equation}
\Psi[\hat{q}^A] = \int \delta q^A
\mathrm{exp}(-I[q^A(\lambda)]/\hbar)
\end{equation}

\noindent where $q^A(\lambda)$ are histories of dynamical variables 
 such as $a, \beta_\pm, etc$, and $I[q^A(\lambda)]$ is their Euclidean action. 
The sum is carried out over cosmological geometries that are regular on a manifold with 
only one boundary at which $q^A(\lambda)$ take some prescribed real values. 
The integration is carried out along a suitable complex 
contour which ensures the convergence of the integral 
and the reality of the result. 

\indent We use the steepest descent approximation to evaluate the path integral. 
We consider only paths that extremize the action.

\begin{equation}
 \frac{\delta I}{\delta q^A(\lambda)} = 0 \label{henbun}
\end{equation}

\noindent then we can write

\begin{equation}
 \Psi(\hat{q}^A) = \sum_{\mathrm{ext}} 
\mathrm{exp[-{\cal A}_\mathrm{ext}/\hbar]}
\end{equation}

\noindent where the summation run over the extrema.
The equations given by (\ref{henbun}) are recognized as Euclidean Einstein equations.
Note that these extremized paths are complex contours, not real 
with a few exceptions (Halliwell and Hartle \cite{Halliwell}).
 Of course the physical quantities are real, thus 
we will set the boundary value $\hat{q}^A$ as real. Therefore
the extremized paths are {\it fuzzy instanton}, which started as
complex Euclidean geometry and growing to asymptotically real Lorentzian at large
scale.\\
\noindent We consider only the lowest order in $\hbar$, thus

\begin{eqnarray}
 {\cal A}_\mathrm{ext} &=& I_\mathrm{ext}(q^A) + O(\hbar) \nonumber\\
 &\equiv& I[q_\mathrm{ext}^A(\lambda)] + O(\hbar)
\end{eqnarray}

\noindent From now we drop the subscript 'ext'.
Since the Euclidean action is complex, we can divide the action into  
real part and imaginary part, $I=I_R-iS$ and write the wave function as follows;
\begin{eqnarray}
 \Psi(\hat{q}^A) &=& A(q^A)e^{iS(q^A)/\hbar} \nonumber \\
 A(q^A) &\equiv& e^{-I_R(q^A)/\hbar}
\end{eqnarray}
\noindent The wave function is said to be the approximate semiclassical form 
when  $S(q^A)/\hbar$ varies rapidly and $A(q^A)$ varies slowly. 
Then the wave function in this region of 
minisuperspace predicts an ensemble of classical trajectories. Therefore

\begin{equation}
 |\nabla_A I_R| \ll |\nabla_A S| \label{CC}
\end{equation}

\noindent is called the {\it classicality condition}. 
Using conserved current 

\begin{equation}
 J_A \equiv -\frac{i\hbar}{2} 
  \left( \Psi^*\frac{\partial \Psi}{\partial q^A}-\Psi\frac{\partial
   \Psi^*}{\partial q^A}  \right)
\end{equation}

\noindent we assume that the relative probability density of classical
history is the component of the conserved current along the specific
surface in minisuperspace. 

\begin{eqnarray}
 \rho(q^A) &\equiv& J \cdot n = |A(q^A)|^2 \nabla_n S(q^A) \nonumber \\
  &=& e^{-2I_R(q^A)}\nabla_nS
\end{eqnarray}

\noindent $n$ is the normal of the spacelike surface in minisuperspace.
If we get the complex paths 
which extremize the Euclidean action and
satisfy the classicality condition, we are able to calculate the classical
Lorentzian history by using these paths as initial condition.

\begin{eqnarray}
 S_{Lorentz}(q^A_L) &=& -\mathrm{Im}[I(\hat{q}^A)] \nonumber\\
 q^A_L &=& \hat{q}^A \\ 
 p^A_L &=& -\mathrm{Im}(p^A) \nonumber
\end{eqnarray}

\noindent Summarize this section\bigskip

\noindent {\it STEP1}\\
 Calculate equation (10) to get the Euclidean Einstein equations.\bigskip

\noindent{\it STEP2}\\
Solve these equations in complex plane under the condition that the boundary values are real.\bigskip

\noindent{\it STEP3}\\
 Check whether the solutions satisfy the classicality
condition (\ref{CC}) or not.\bigskip

\noindent{\it STEP4}\\
Solve the classical Lorentzian paths and calculate the relative probabilities of these
histories.


\section{NO-BOUNDARY WAVE FUNCTION FOR MIXMASTER UNIVERSES}

In this section we calculate the steepest descent approximation to the no-boundary wave function for the mixmaster universes 
according to the above prescriptions. We include the cosmological constant and a
 minimally coupled single scalar field with quadratic potential.

By replacing $t\to-i\lambda, \, iS\to-I$ in (\ref{Laction}),
the Euclidean action becomes

\begin{equation}
 I[a(\lambda),\Phi(\lambda),\beta_\pm(\lambda)] = I_\mathrm{IX}+I_\Phi
\end{equation}

\noindent where $I_\mathrm{IX}$ is the action of spacetime

\begin{equation}
 I_\mathrm{IX} =
 \frac{\pi}{G}\int d\lambda 
Ne^{3\alpha} \left[\frac{6}{N^2} \left (-\alpha'^2 + 
\beta_+'^2 + \beta_-'^2\right) - {}^3R_\mathrm{IX} + 2\Lambda \right]
\end{equation}

\noindent with $ A'=dA/d\lambda$. $I_\Phi$ is the action for the minimally coupled scalar field with a 
quadratic potential

\begin{eqnarray}
I_\Phi &=& \int d^4x \sqrt{-g} {\cal L}_{matter} \nonumber\\
&=& (4\pi)^2 \int d\lambda Ne^{3\alpha} \left[\frac{1}{2N^2}\Phi'^2
-\frac{1}{2}m^2\Phi^2 \right]
\end{eqnarray}

\noindent We have assumed that scalar field is homogeneous, 
 and  $m$ is the mass of scalar field. 
For later convenience, we rescale the variables as follows;

\begin{eqnarray}
 e^\alpha &=& a \\
 N &\to& \sqrt{\frac{3}{\Lambda}}N \\
 a &\to& \frac{1}{2}\sqrt{\frac{3}{\Lambda}}a \\
 \Phi &=& \sqrt{\frac{3}{4\pi G}}\phi \\
 m    &=& \sqrt{\frac{\Lambda}{3}}\mu 
\end{eqnarray}

\noindent Then the explicit form of the Euclidean action takes the following form

\begin{multline}
I[a(\lambda),\Phi(\lambda),\beta_\pm(\lambda)]\\
 = \frac{9\pi}{4\Lambda G} \int_{0}^{1} d\lambda Na^3
 \left[ \frac{1}{N^2} \left( -\frac{a'^2}{a^2}
 + \beta_+'^2 + \beta_-'^2 \right) \right.\\
 \left. + 1 - \frac{1}{a^2}(1-V(\beta_\pm))
 + \frac{\phi'^2}{N^2}+\mu^2\phi^2 \right]
\end{multline}

\noindent where

\begin{multline}
 V(\beta_\pm) = 1
-\frac{4}{3}e^{-2\beta_+}\cosh{2\sqrt{3}\beta_-}\\
+\frac{1}{3}e^{-8\beta_+}
-\frac{2}{3}e^{4\beta_+}(1-\cosh{4\sqrt{3}\beta_-})
\end{multline}

\noindent Note that the variables are all complex valued.  
Therefore the parameter $\lambda$ can be chosen to be real, and 
we choose $\lambda=0$ at the starting point and $\lambda=1$ is our boundary.

Taking the variation to the above Euclidean action and using the new complex parameter
$d\tau=Nd\lambda$, we get five equations

\begin{align}
& \dot{a}^2-a^2(\dot{\beta_+}^2+\dot{\beta_-}^2) \nonumber\\
&\hspace{2em}+a^2-(1-V(\beta_\pm))
+ a^2\dot{\phi}^2  +a^2\mu^2\phi^2 = 0\\
&\ddot{a}+\frac{\dot{a}^2}{2a}+\frac{3}{2}a(\dot{\beta_+}^2+\dot{\beta_-}^2)
\nonumber\\
&\hspace{3em}-\frac{1}{2a}(1-V(\beta_\pm))
+\frac{3}{2}a(\dot{\phi}^2+\mu^2\phi^2)=0\\
&\ddot{\phi}+3\frac{\dot{a}}{a}\dot{\phi}-\mu^2\phi=0\\
&\ddot{\beta_\pm}+3\frac{\dot{a}}{a}\dot{\beta_\pm}
-\frac{1}{2a^2}\frac{\partial V(\beta_\pm)}{\partial\beta_\pm}=0
\end{align}

\noindent where $\dot{A}=dA/{d\tau}$. 
Four in the above five equations are independent. Finally, the Euclidean action is written as

\begin{equation}
\begin{split}
I[&a(\tau),\phi(\tau),\beta_\pm(\tau)]\\
 &=\frac{9\pi}{4\Lambda G}\int_{C(0,v)} d\tau \left[
a \dot{a}^2+a^3(\dot{\beta_+}^2+\dot{\beta_-}^2)
\right.\\
& \left.\hspace{4em}+a^3 - a(1-V(\beta_\pm)) 
+a^3\dot{\phi}^2+\mu^2 a^3 \phi^2 \right]\\
&= \frac{9\pi}{2\Lambda G}\int_{C(0,v)} d\tau
a[a^2(1+\mu^2\phi^2)-1+V(\beta_\pm)]
\end{split}
\end{equation}

\noindent Note that the action is given by line integral over the contour
in complex $\tau$-plane, $C_{(0,v)}$, start from origin to $v=X+iY$.
Following Lyons \cite{Lyons}, we integrate along the real axis to $X$ first, and then along
the imaginary axis to $v=X+iY$ where the variables are asymptotic real.
 
At the initial position $\tau=0$, we require that the scale factor $a(0)=0$ and
all other variables are regular. Thus we can Taylor expand the variables near the origin $\tau=0$ and get 
the following expressions
\begin{align}
&a(\tau) = \tau - \frac{1}{6} (1+\mu^2 \bar{\phi_0}^2) \tau^3 + O(\tau^5)\\
&\phi(\tau) = \bar{\phi_0}+\frac{1}{8}\mu^2\bar{\phi_0}\tau \nonumber\\
&\hspace{4em}+
 \frac{1}{48}\mu^2\bar{\phi_0}(2+\mu^2+2\mu^2\bar{\phi_0}^2)\tau^3
 +O(\tau^5)\\
&\beta_+(\tau) = \bar{P_2}\tau^2 \nonumber\\
&\hspace{3em} + \left[ \frac{7}{24}(1+\mu^2 \bar{\phi_0}^2)\bar{P_2}
-10\bar{P_2}^2+10\bar{M_2}^2 \right]\tau^4 \nonumber\\
&\hspace{7em}+ O(\tau^6)\\
&\beta_-(\tau) = \bar{M_2}\tau^2 \nonumber\\
&\hspace{3em}+ \left[ \frac{7}{24}(1+\mu^2\bar{\phi_0}^2)\bar{M_2}
+5\bar{P_2}\bar{M_2} \right]\tau^4 +O(\tau^6)
\end{align}

\noindent This shows that we can choose three complex numbers  
$\bar{\phi_0},\bar{P2},\bar{M2}$ as our free free parameters. 
We write them as follows:

\begin{eqnarray}
\bar{\phi_0} &=& |\bar{\phi_0}|e^{i\theta} 
\equiv  \phi_0 e^{i\theta}\\
\bar{P_2} &=& |\bar{P_2}e^{i\Theta_p}
\equiv P_2 e^{i\Theta_p}\\
\bar{M_2} &=& |\bar{M_2}|e^{i\Theta_m}
\equiv M_2 e^{i\Theta_m}
\end{eqnarray}
 

\section{RESULTS}

\subsection{COMPLEX SOLUTION}

\begin{figure*}[p]
 \begin{center}
  \begin{tabular}{cc}
 \resizebox{.4\textwidth}{!}{\includegraphics[scale=0.15,angle=270]{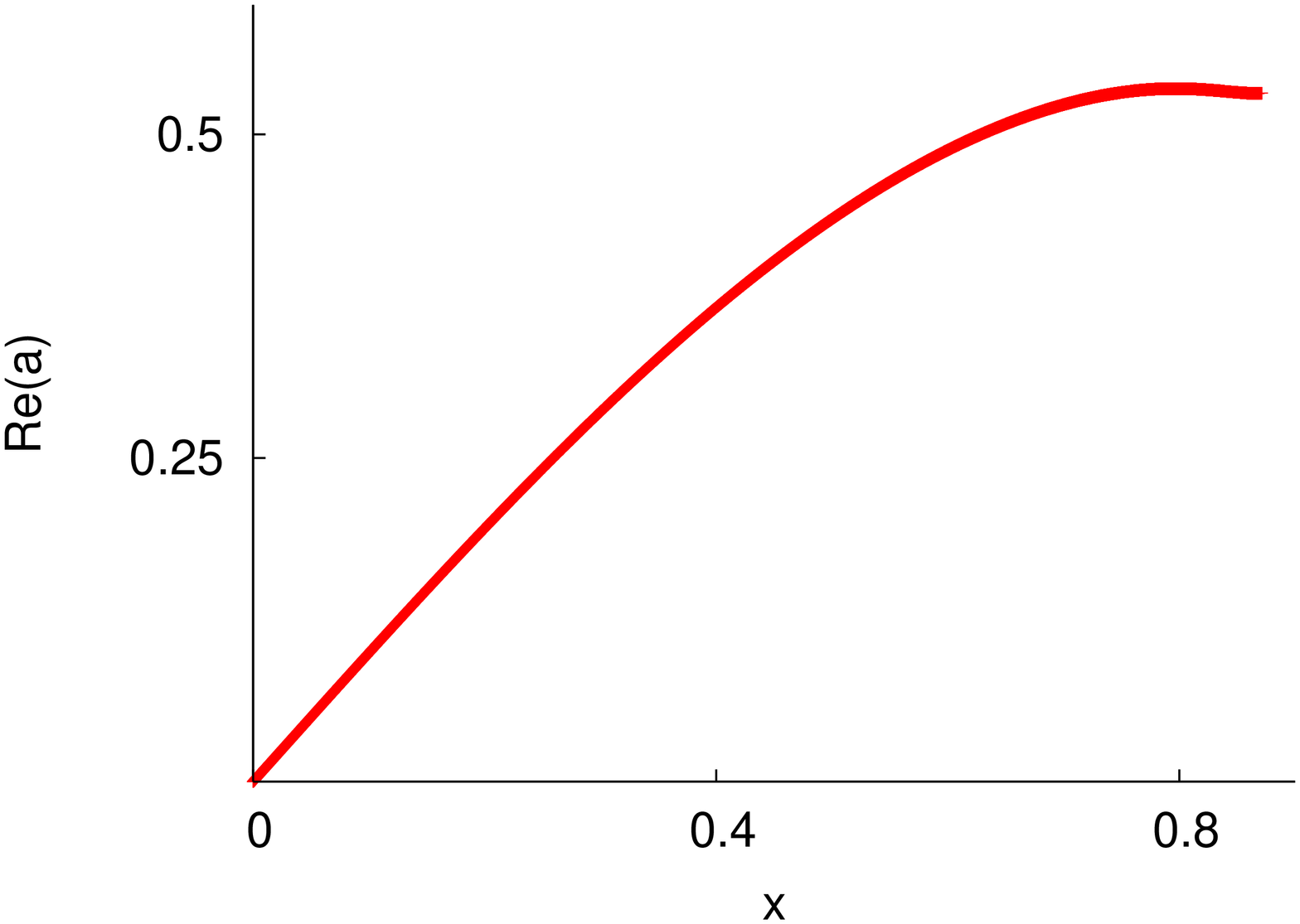}}& 
 \resizebox{.4\textwidth}{!}{\includegraphics[scale=0.15,angle=270]{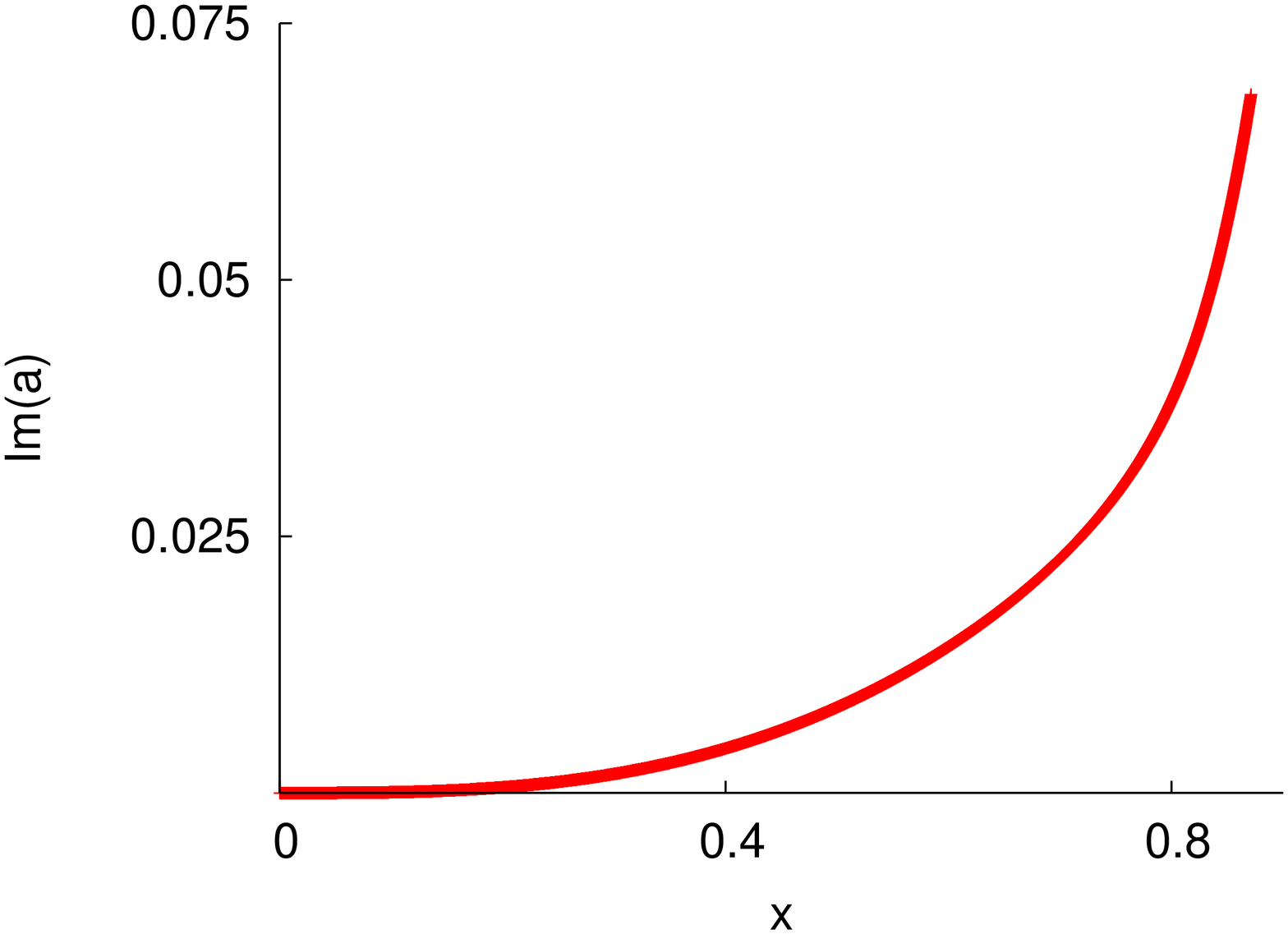}}\\
 \resizebox{.4\textwidth}{!}{\includegraphics[scale=0.15,angle=270]{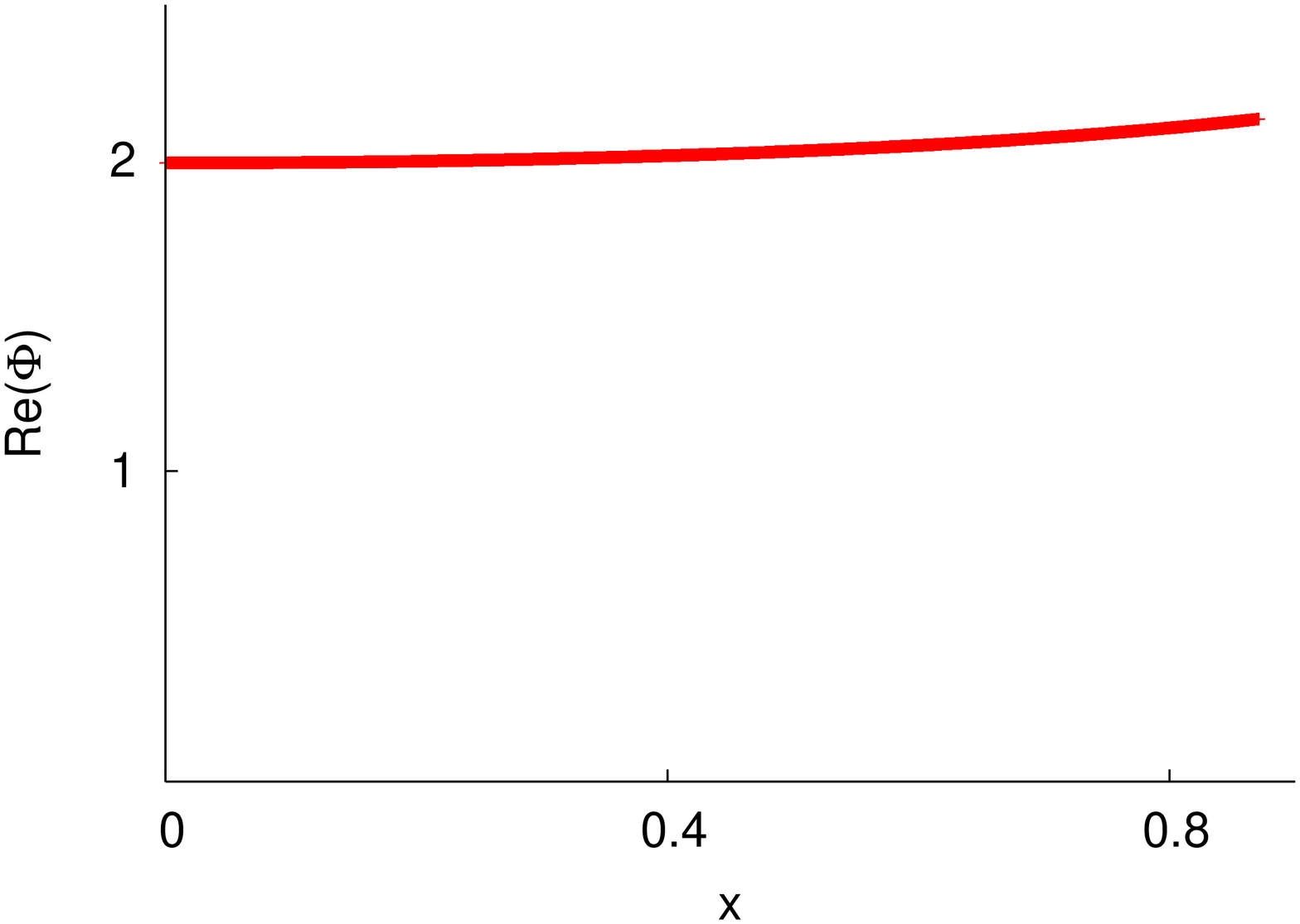}}&
 \resizebox{.4\textwidth}{!}{\includegraphics[scale=0.15,angle=270]{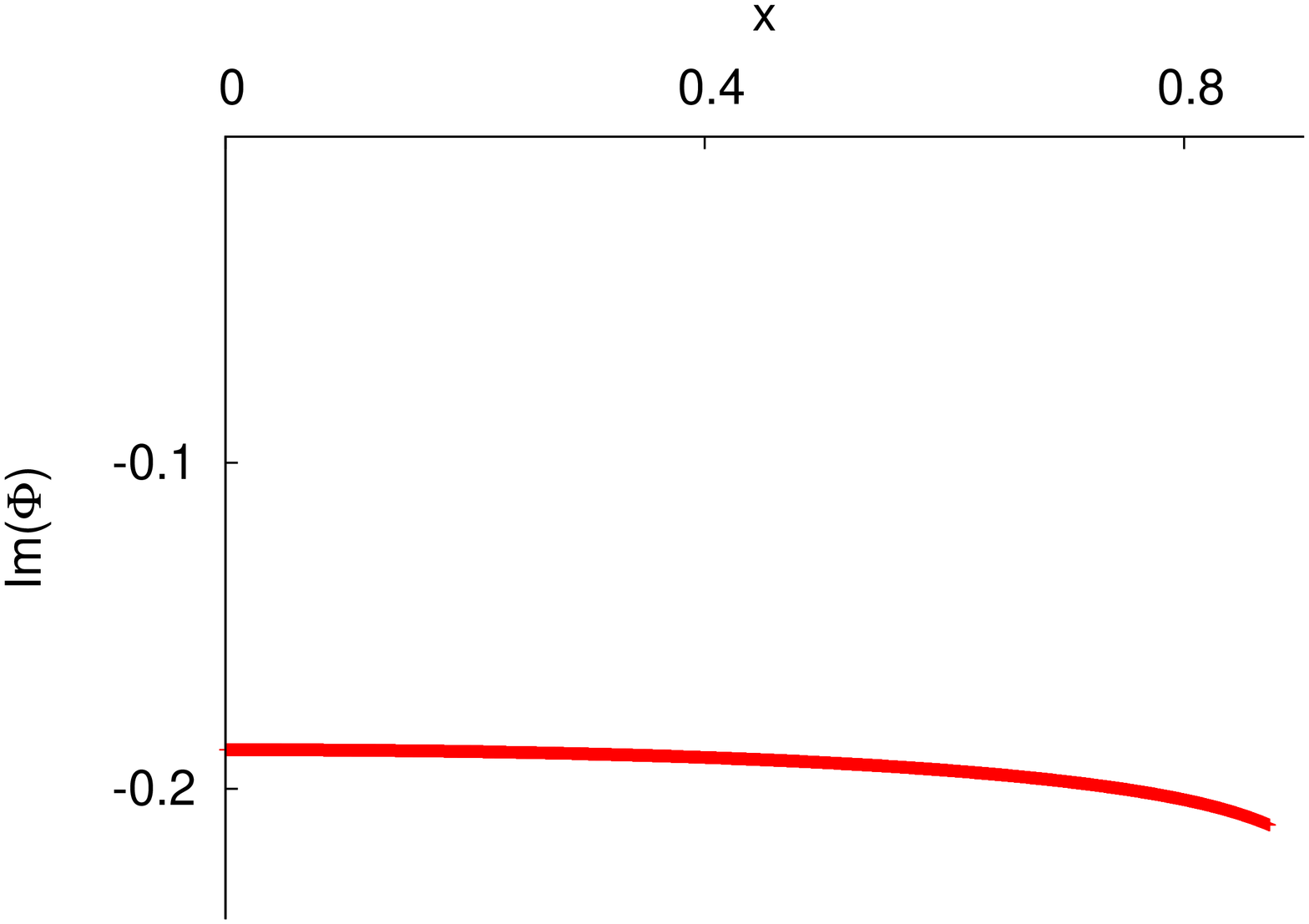}}\\
 \resizebox{.4\textwidth}{!}{\includegraphics[scale=0.15,angle=270]{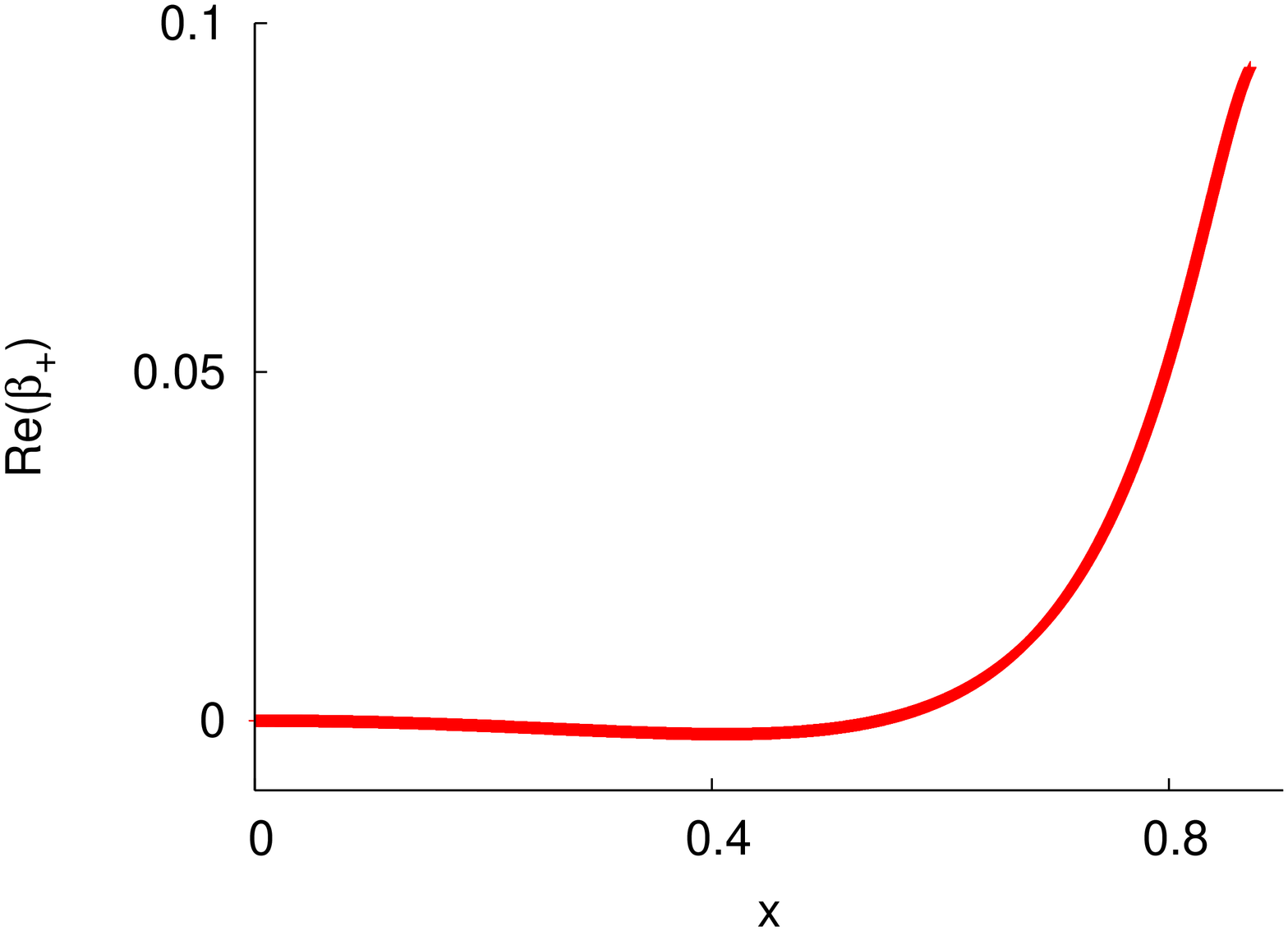}}& 
 \resizebox{.4\textwidth}{!}{\includegraphics[scale=0.15,angle=270]{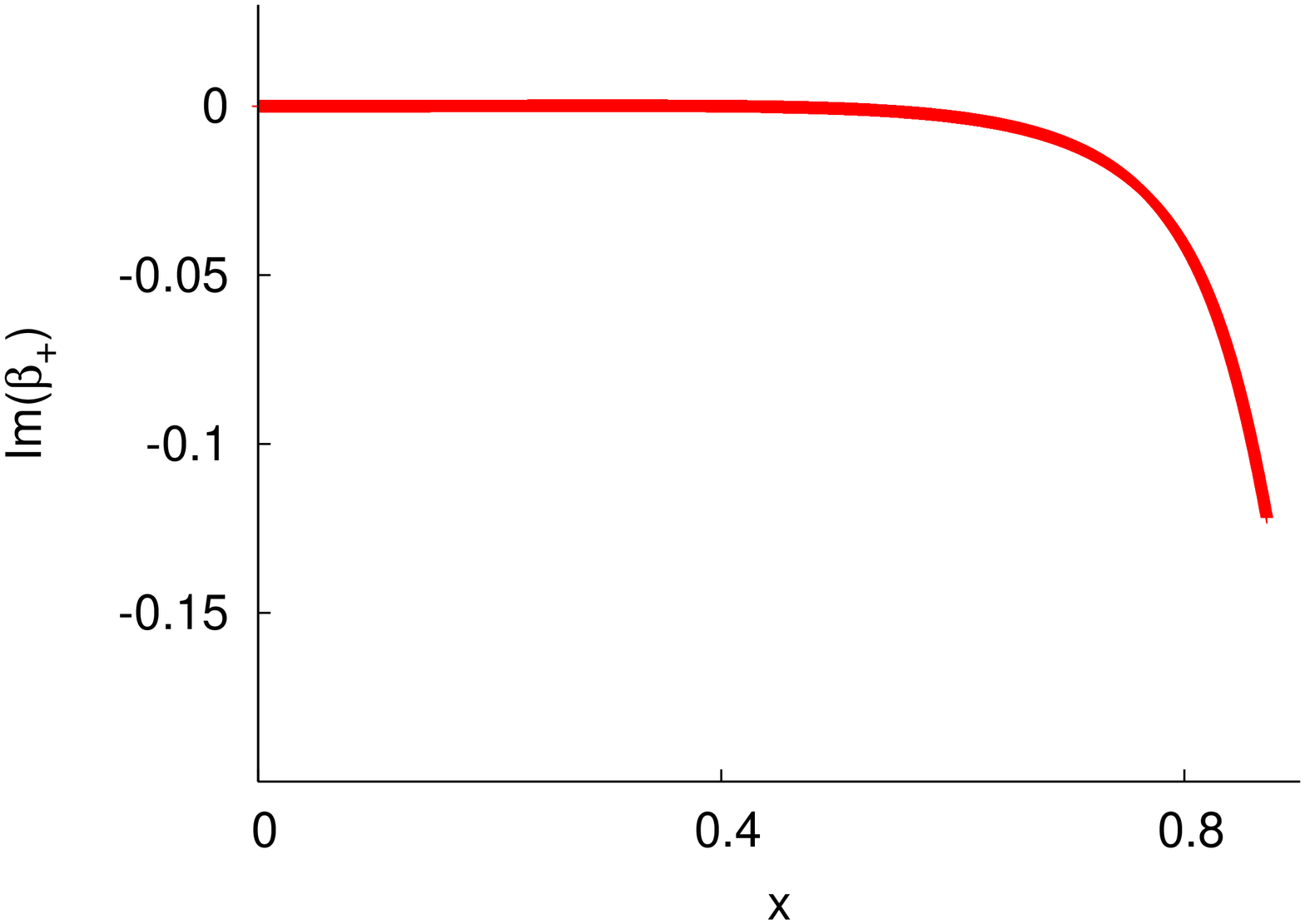}}\\
 \resizebox{.4\textwidth}{!}{\includegraphics[scale=0.15,angle=270]{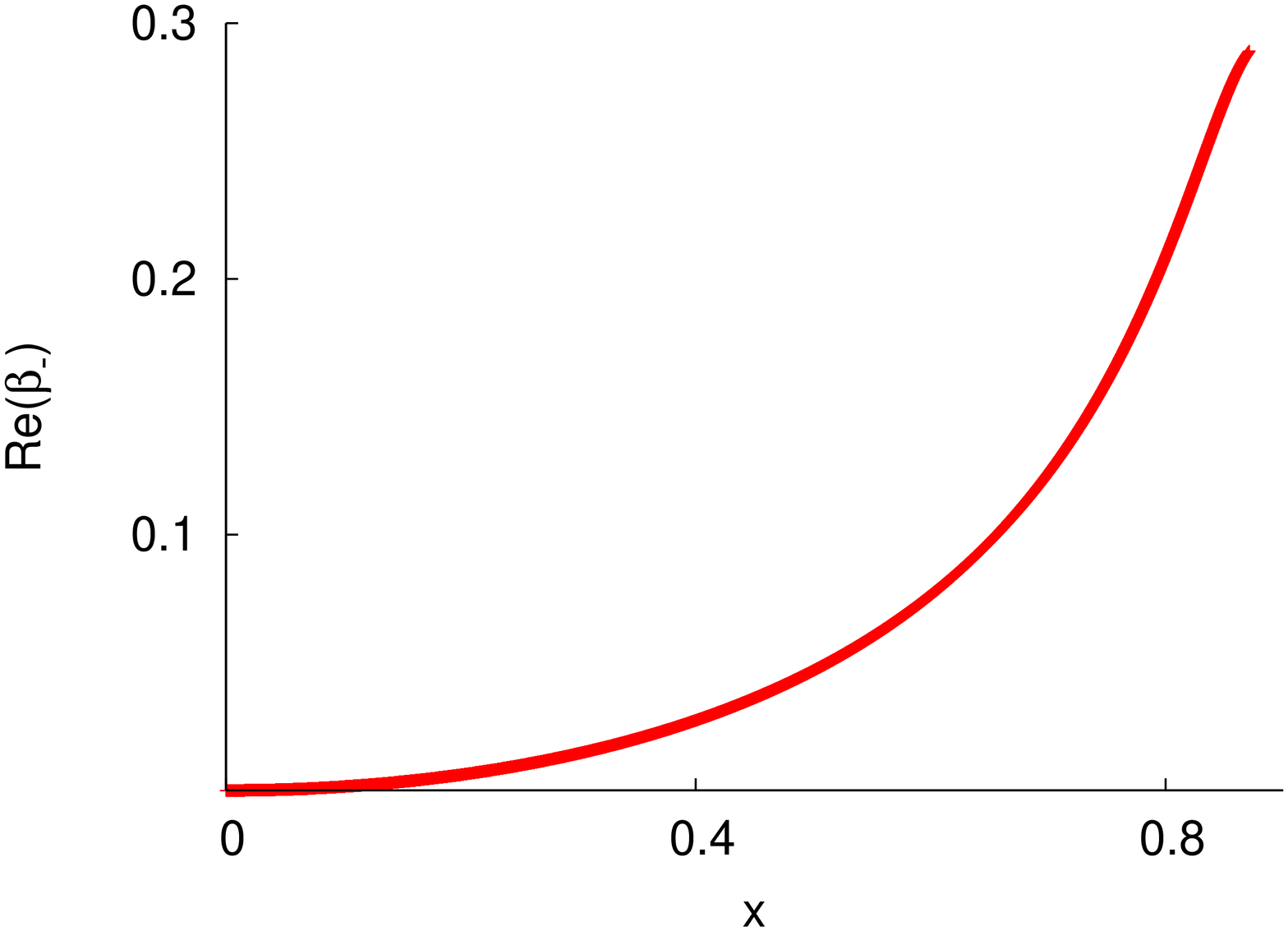}}&
 \resizebox{.4\textwidth}{!}{\includegraphics[scale=0.15,angle=270]{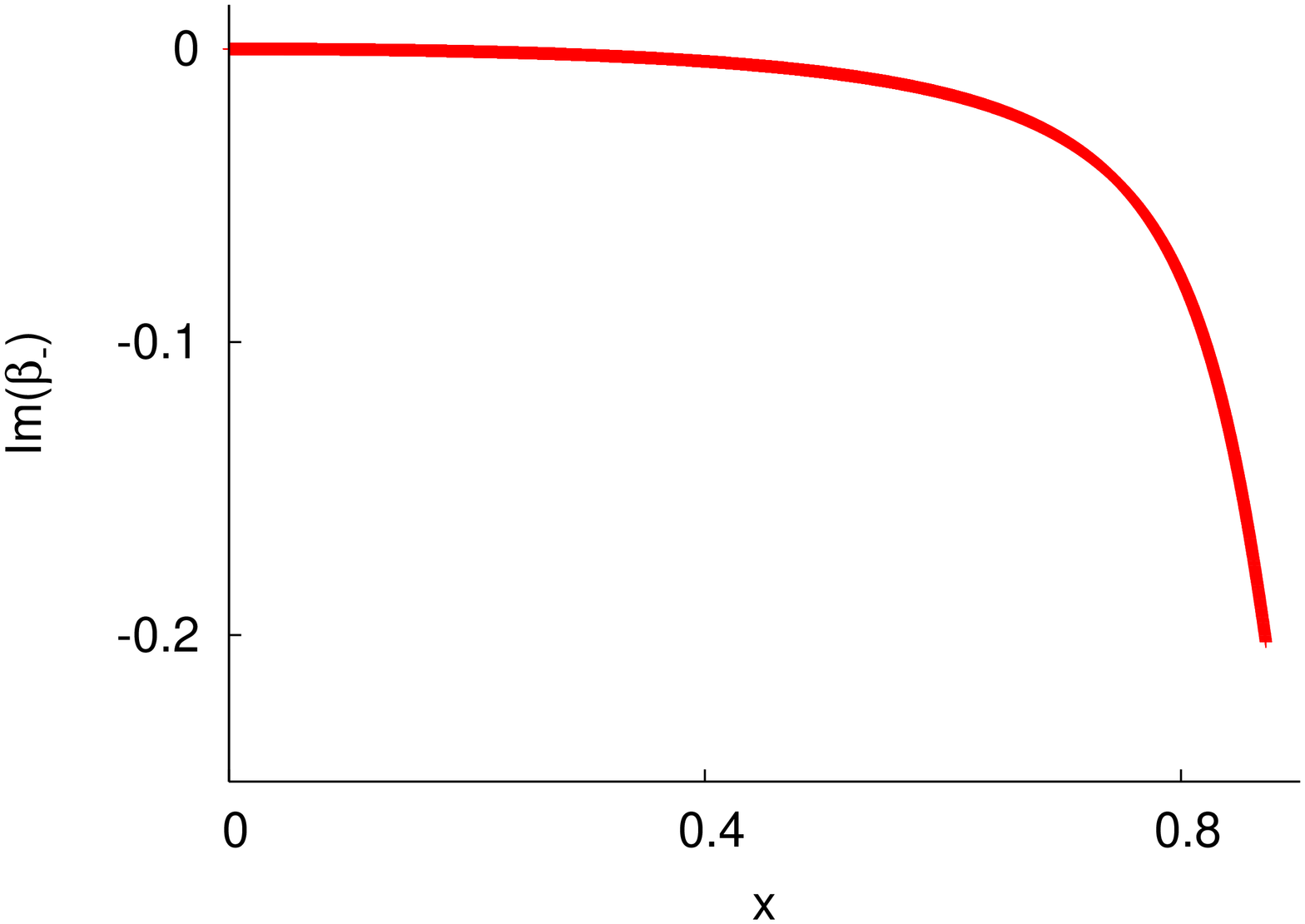}}\\
   \end{tabular}
  \caption{The real and imaginary part of
  the complex-Euclidean solution on X-axis.\newline \hspace{3.4em}
  From top to bottom, scale factor $a$, scalar field $\phi$,
  anisotropy $\beta_+$ and $\beta_-$.\newline \hspace{3.4em}
  Left row are real part, right row are imaginary part of the variables.
  \newline \hspace{3.4em}
  $\mu=3/4, \phi_0=2, P2=0.02, M2=0.15$ 
  and $X=0.871, \theta=0.0942, \Theta_p=3.01, \Theta_m=-0.133$
   \label{fig:EX}}
 \end{center}
\end{figure*}

\begin{figure*}[p]
 \begin{center}
  \begin{tabular}{cc}
 \resizebox{.4\textwidth}{!}{\includegraphics[scale=0.15,angle=270]{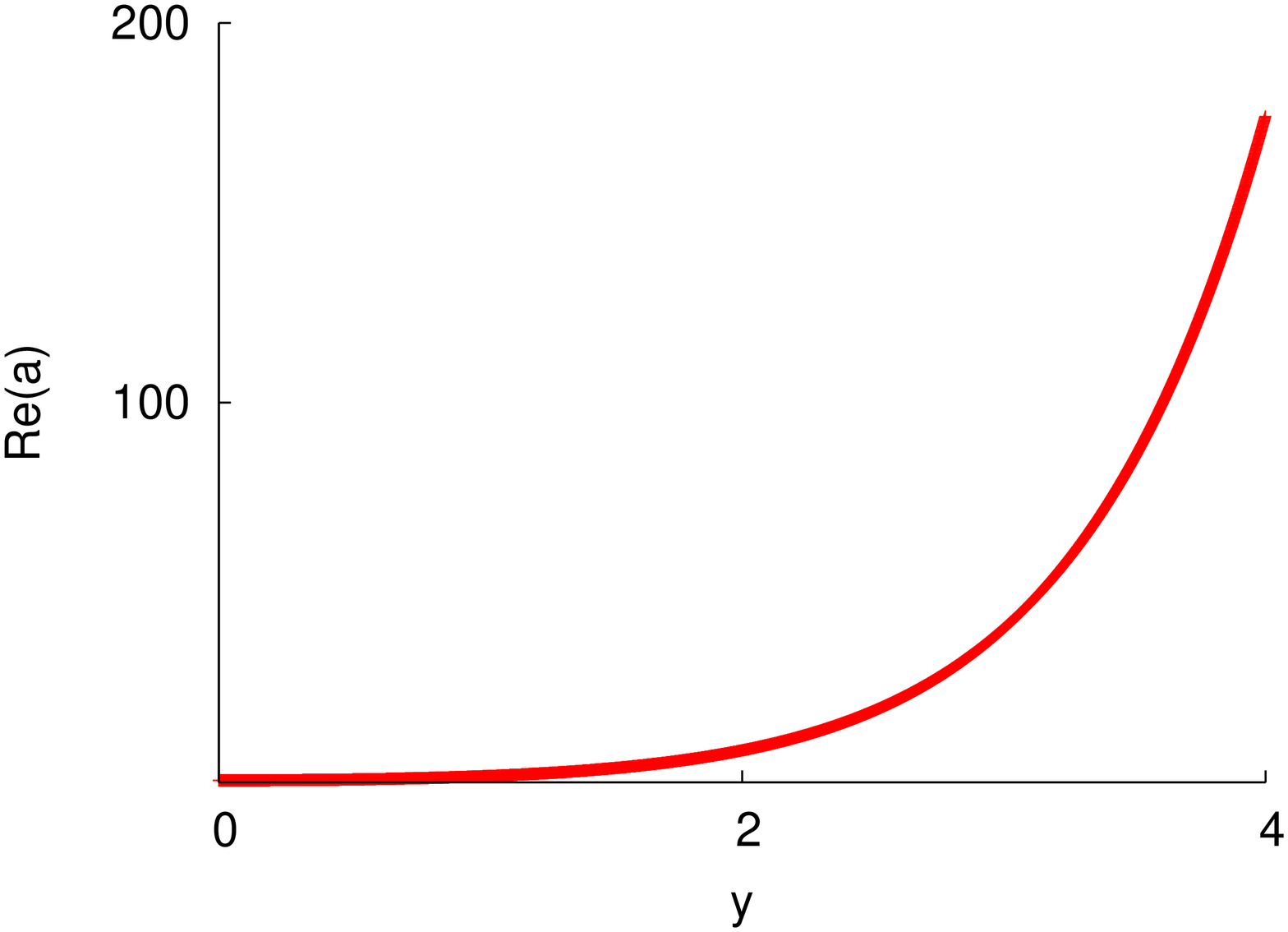}}& 
 \resizebox{.4\textwidth}{!}{\includegraphics[scale=0.15,angle=270]{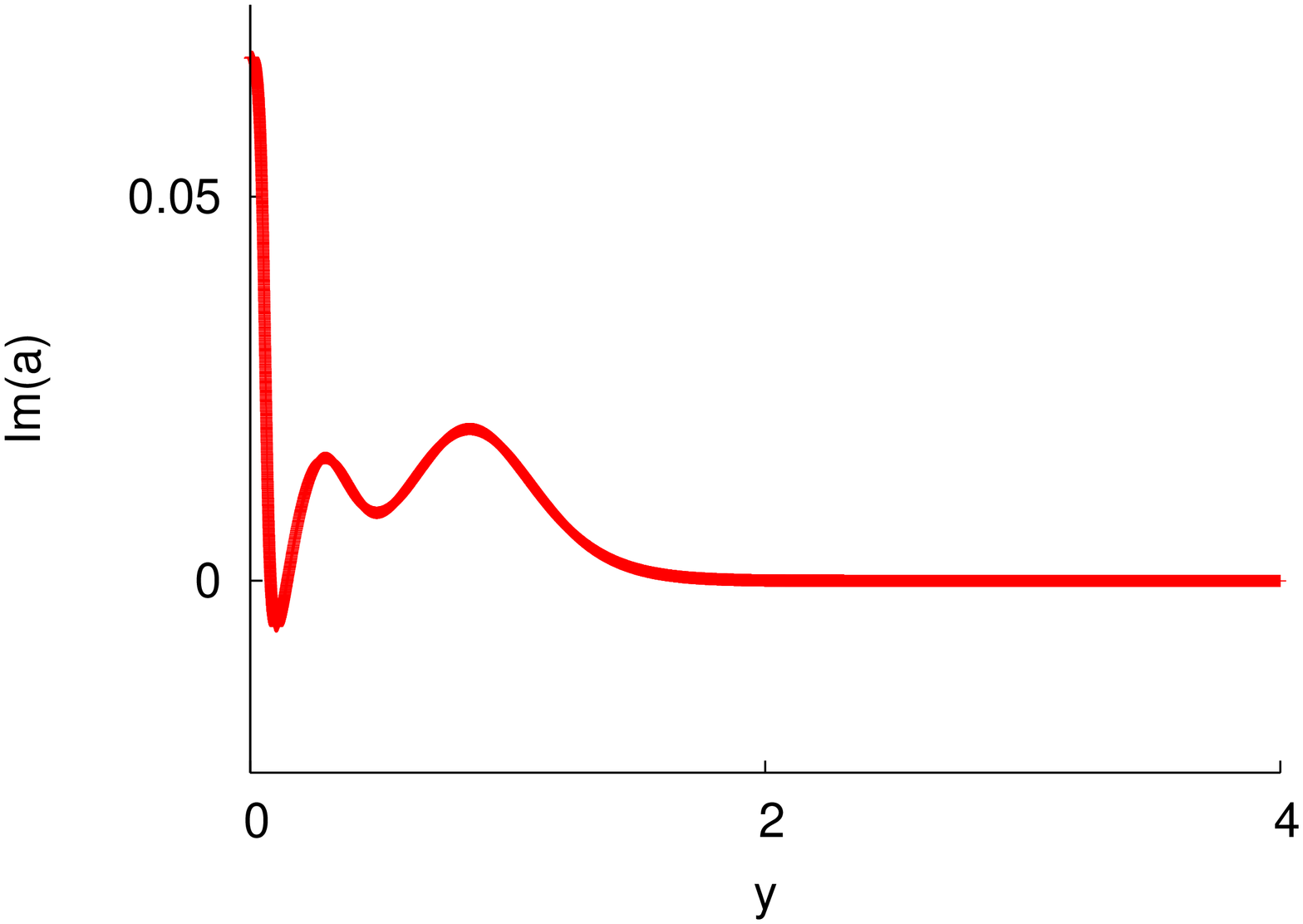}}\\
 \resizebox{.4\textwidth}{!}{\includegraphics[scale=0.15,angle=270]{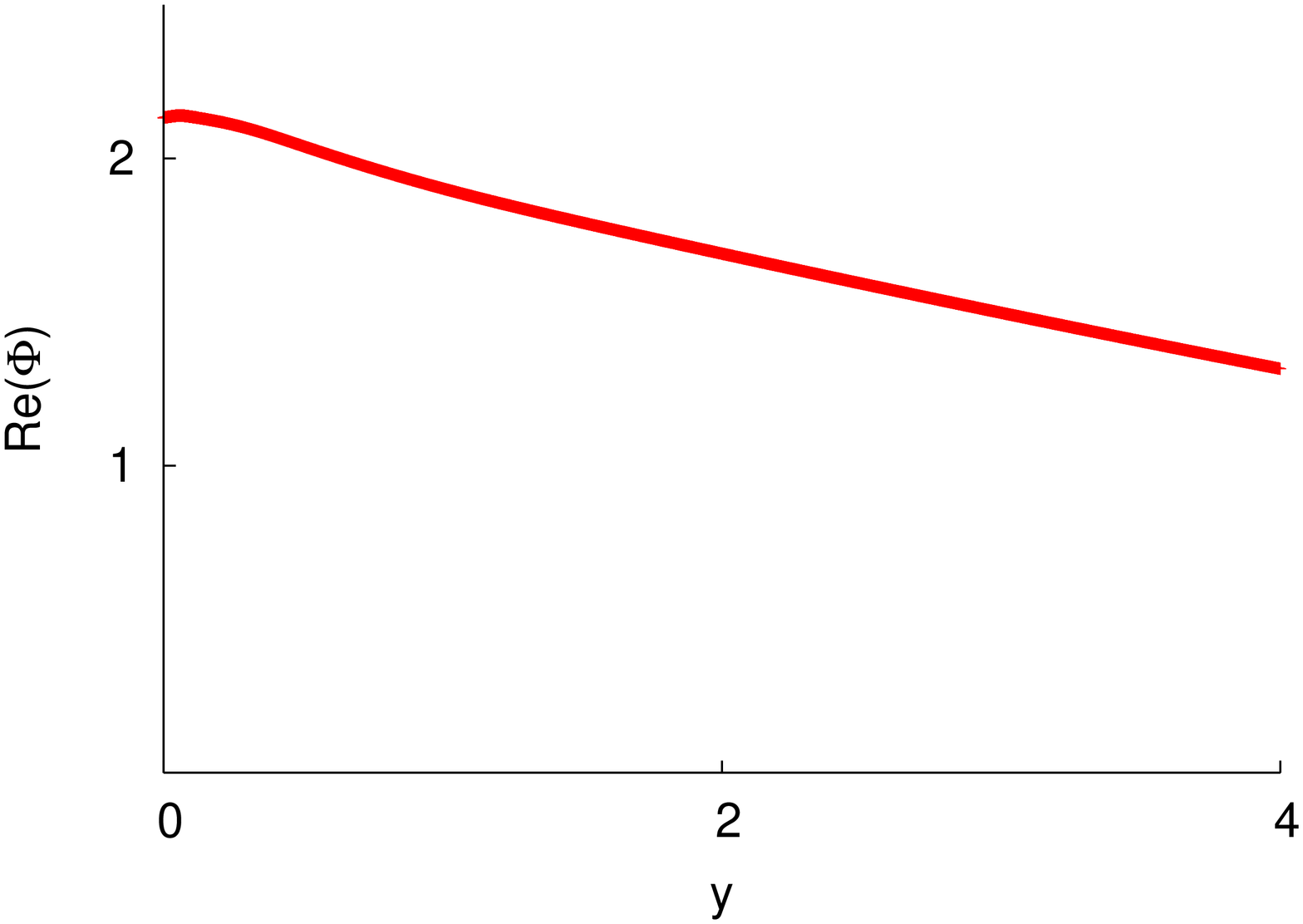}}&
 \resizebox{.4\textwidth}{!}{\includegraphics[scale=0.15,angle=270]{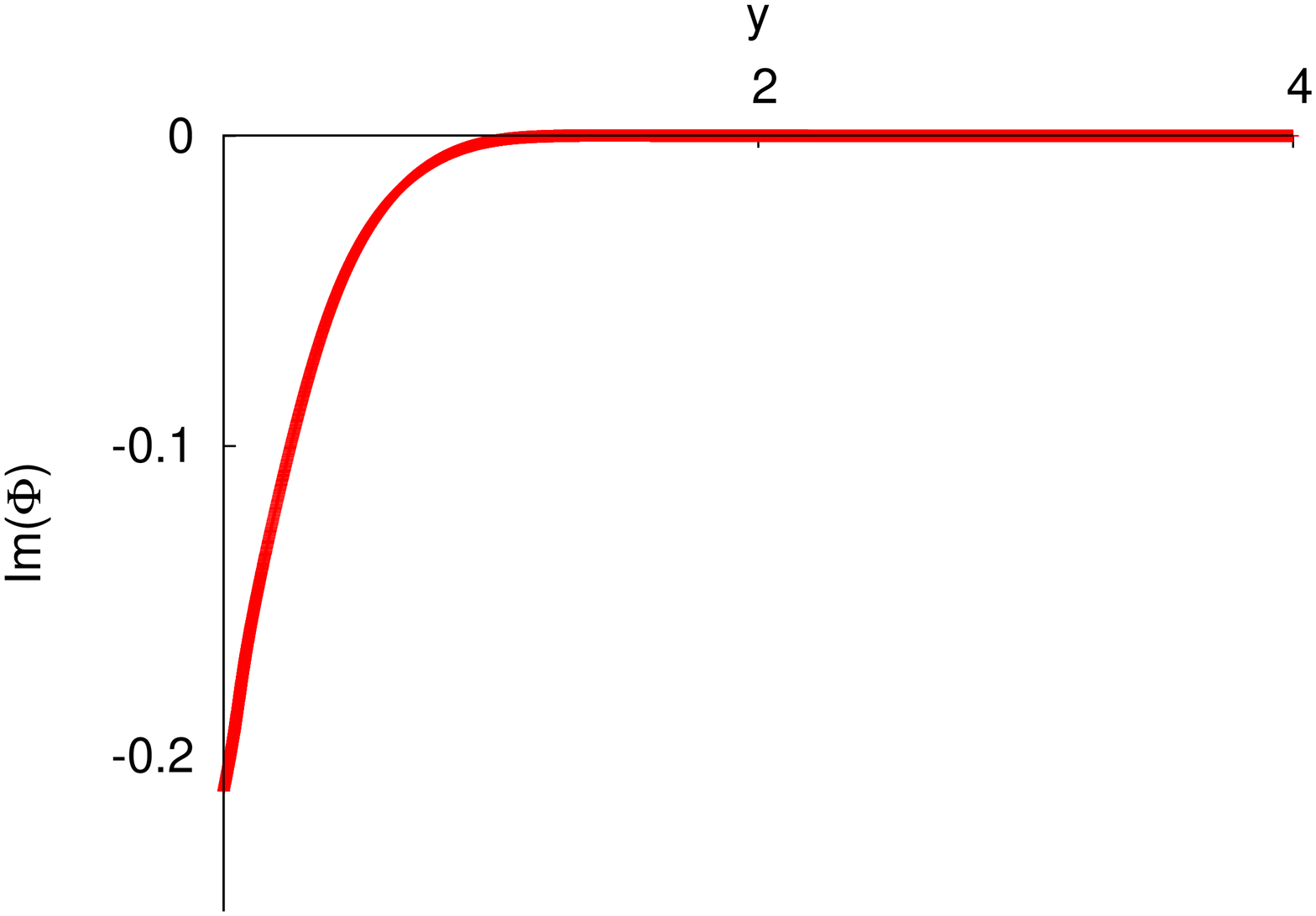}}\\
 \resizebox{.4\textwidth}{!}{\includegraphics[scale=0.15,angle=270]{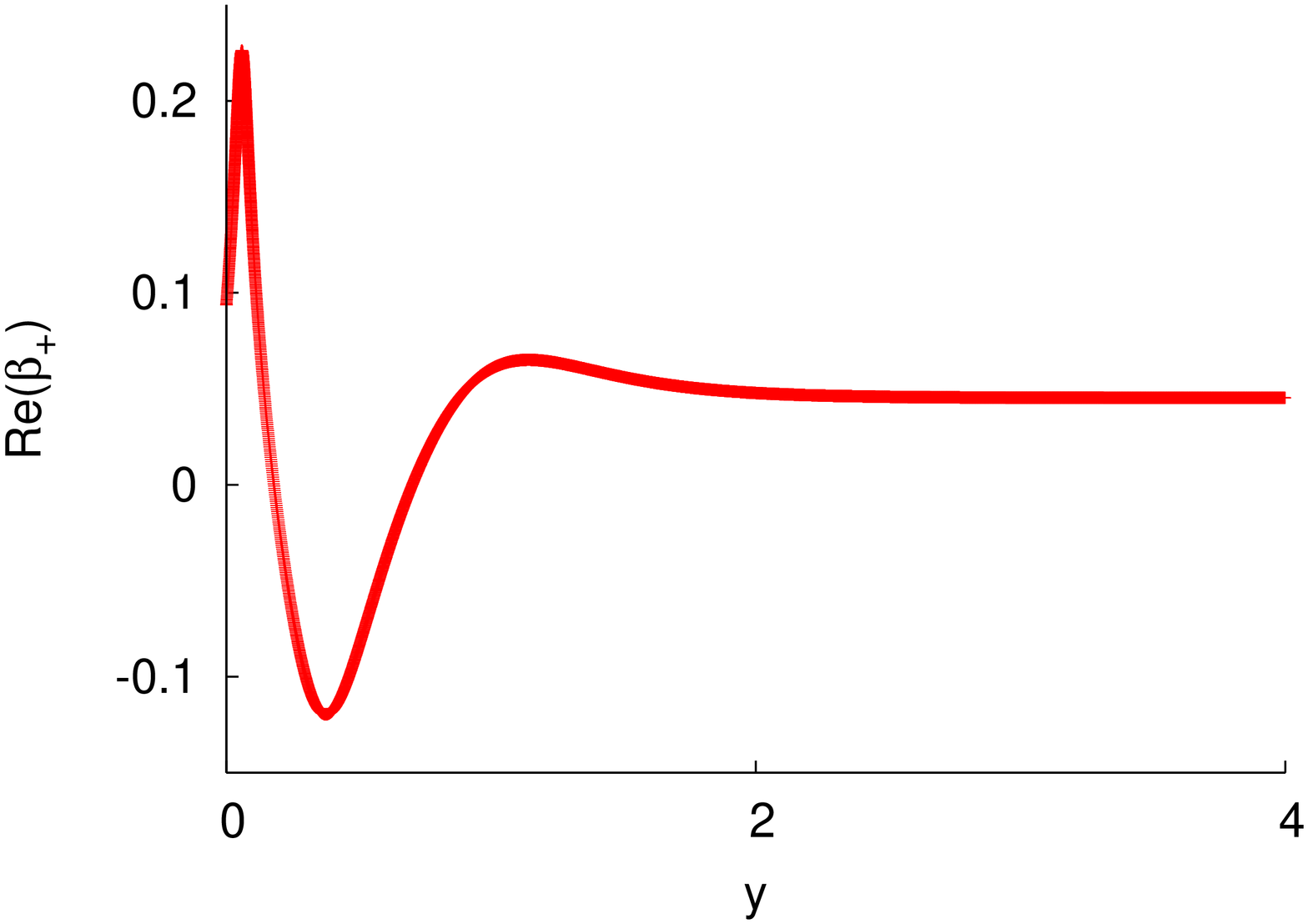}}& 
 \resizebox{.4\textwidth}{!}{\includegraphics[scale=0.15,angle=270]{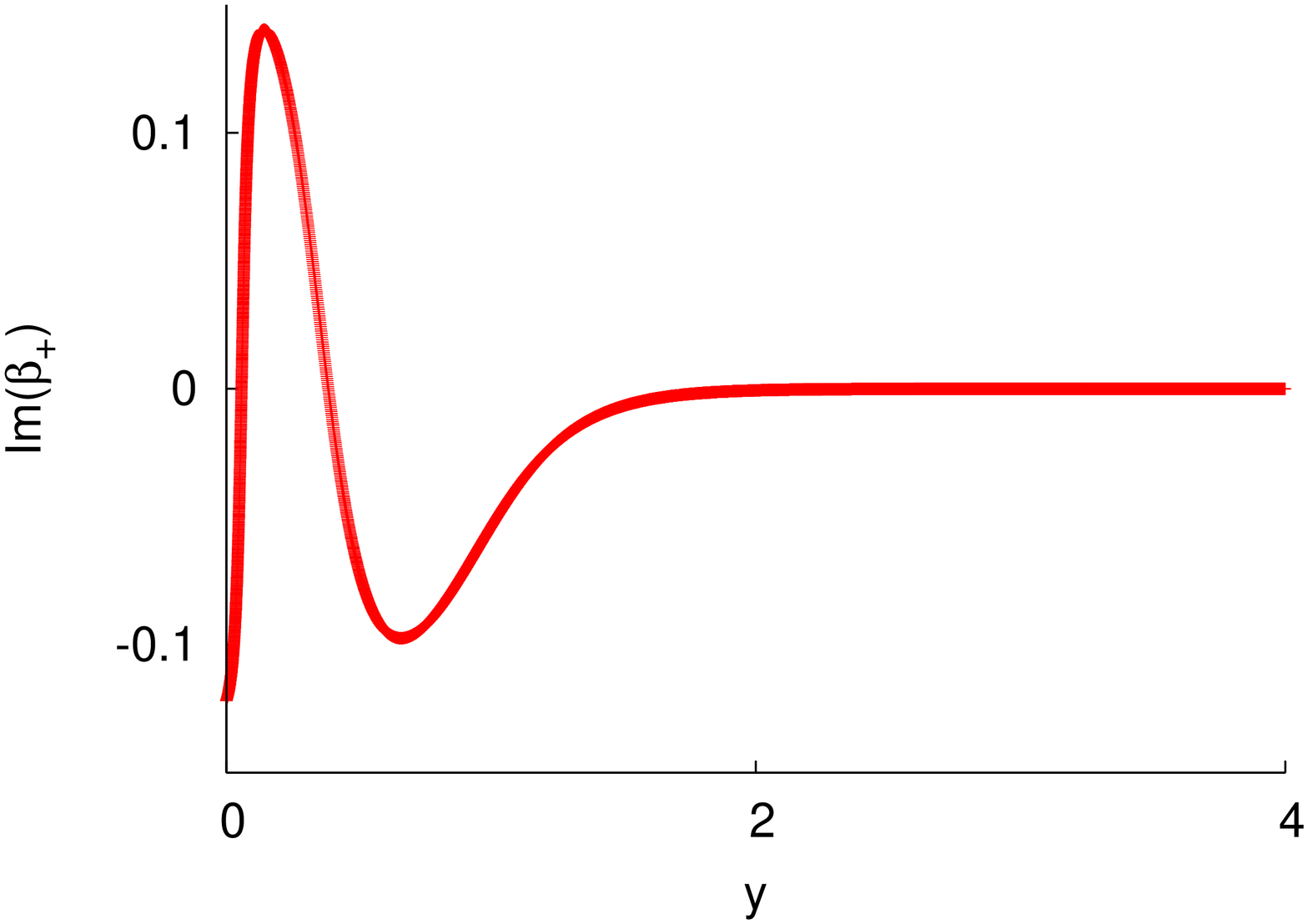}}\\
 \resizebox{.4\textwidth}{!}{\includegraphics[scale=0.15,angle=270]{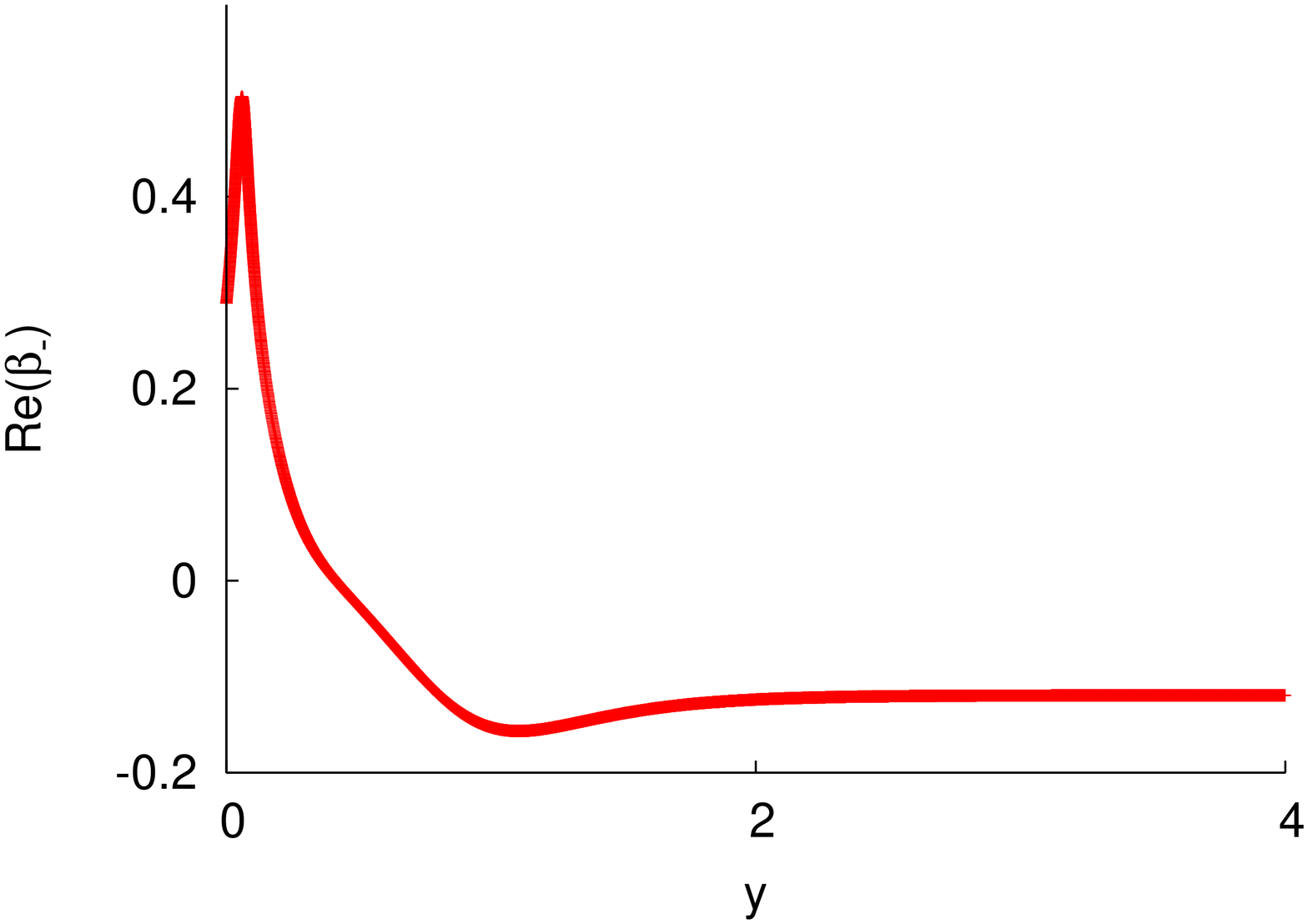}}&
 \resizebox{.4\textwidth}{!}{\includegraphics[scale=0.15,angle=270]{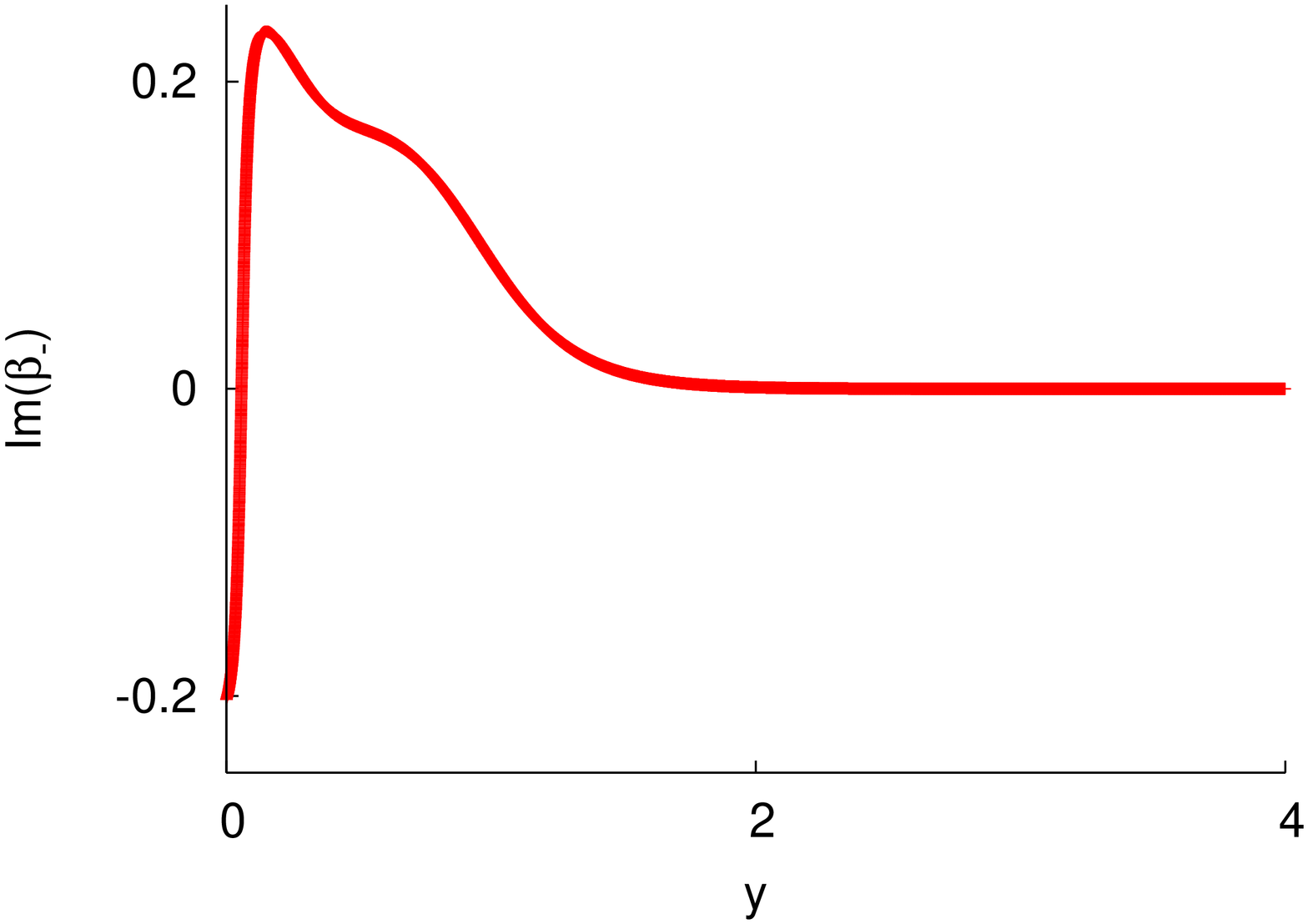}}\\
   \end{tabular}
  \caption{The real and imaginary part of
   the complex-Euclidean solution on Y-axis\newline\hspace{3.4em}
  All imaginary parts become zero at large scale.
  \label{fig:EY}}
  \end{center}
\end{figure*}

We now solve the complex Einstein's equations with the condition that 
the imaginary part of the dynamical variables become zero at large $\tau$.\\
\indent Figure.\ref{fig:EX} and Figure.\ref{fig:EY} show 
the real and imaginary part of the scale factor $a$,
scalar field $\phi$, anisotropy $\beta_{\pm}$.
The parameters are $\mu = 3/4$, $\phi_{0} = 2$, $P_2 = 0.02$,
and $M_2 = 0.15$. If $P_2=M_2=0$, the parameters are same as
HHH's model. All imaginary parts become zero at large scale as seen by these figures.

\begin{figure*}[tbp]
 \begin{center}
  \includegraphics[scale=0.5,angle=270]{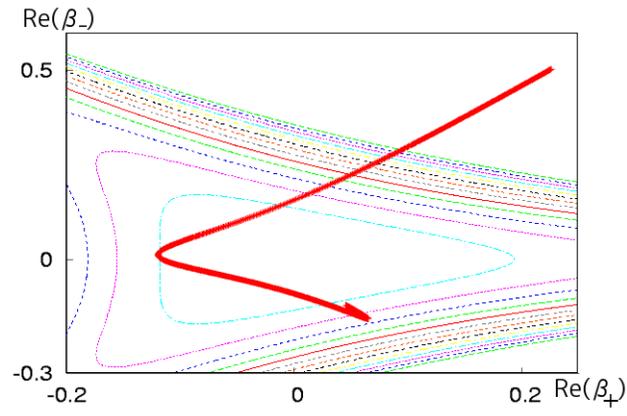}
  \end{center}
 \caption{The real part of $\beta_+$ vs $\beta_-$.\newline\hspace{3.4em}
 The universe point comes from high anisotropy to low anisotropy
 with some bounce.\newline\hspace{3.4em}
 Note that this is in Euclidean solution, not classical Lorentzian.
 \label{fig:Ebeta}}
\end{figure*}

\begin{figure*}[tbp]
 \begin{center}
  \begin{tabular}{cc}
 \resizebox{.49\textwidth}{!}
   {\includegraphics[scale=0.2,angle=270]{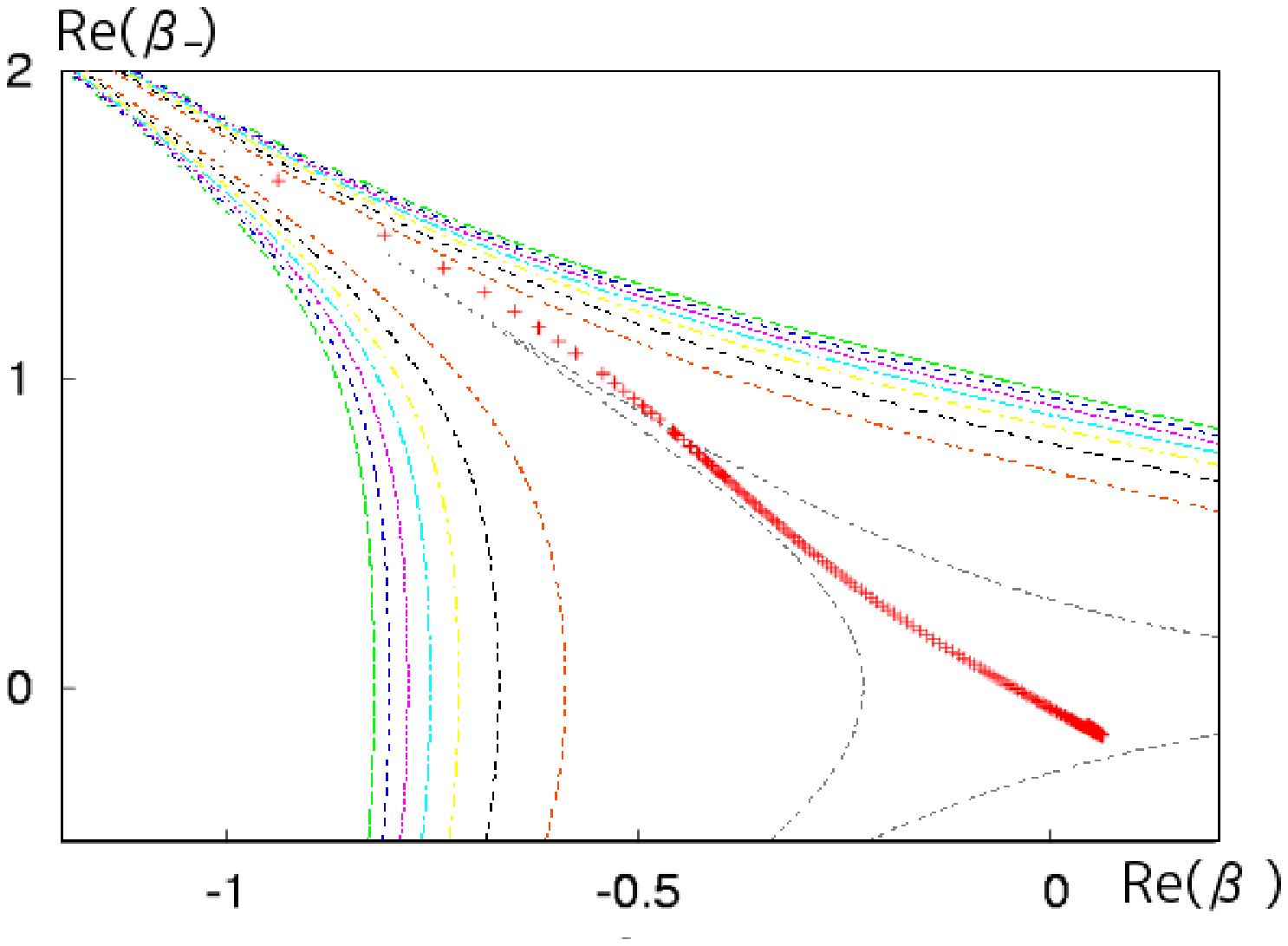}}& 
 \resizebox{.49\textwidth}{!}
   {\includegraphics[scale=0.2,angle=270]{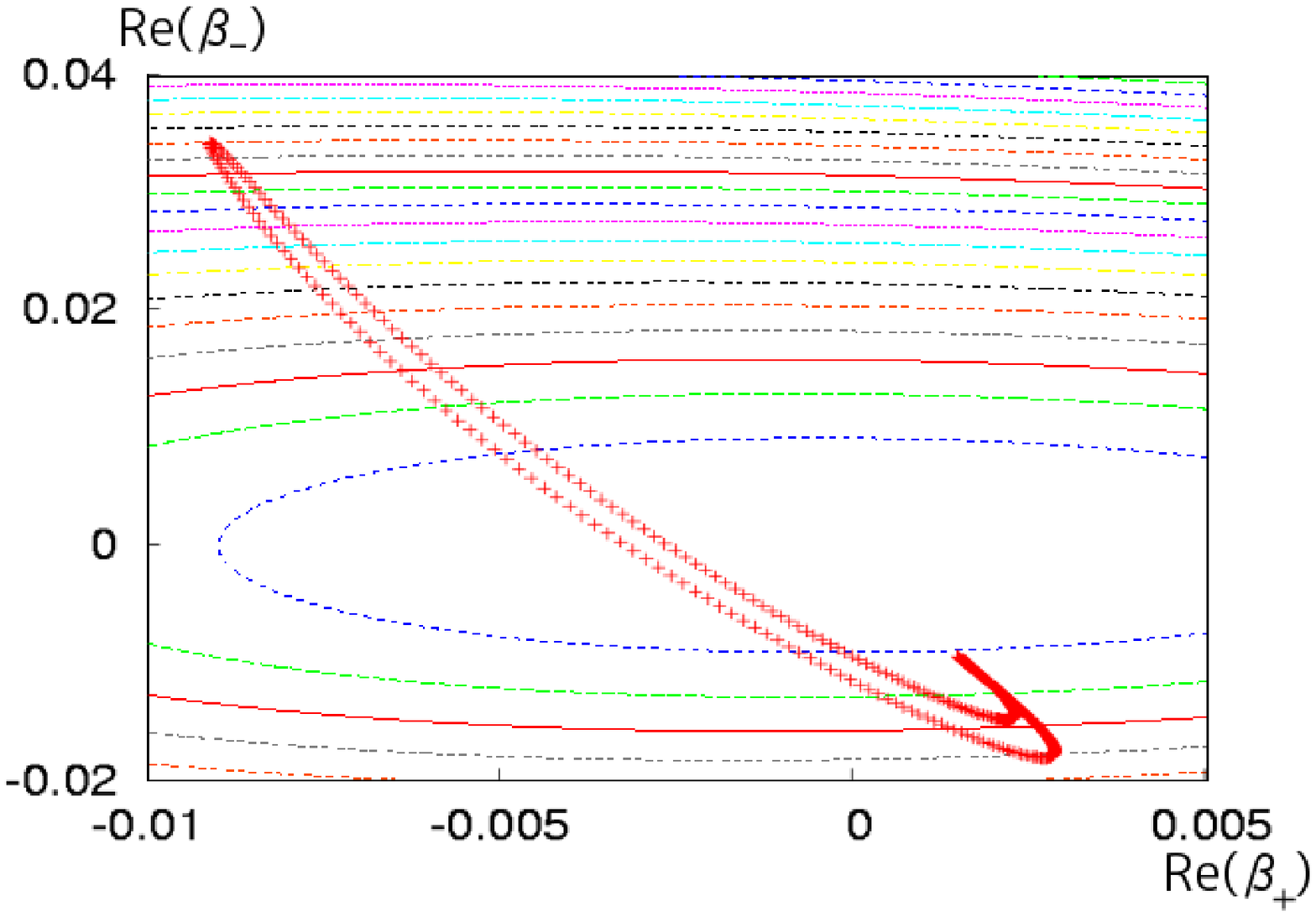}}
  \end{tabular}
  \caption{$\beta_+$ vs $\beta_-$ in classical history.
  Left is model A, right is model B.
  \label{fig:Lbeta}}
 \end{center}
\end{figure*}

Figure.\ref{fig:Ebeta} shows the real part of $\beta_{+}$ vs $\beta_{-}$ in the triangular-shaped potential.
We can see that the universe point bounces between steep walls of the potential.
However, note that this is an Euclidean solution and not directly corresponds to the classical history. 
In fact, the corresponding Lorentzian history has completely different trajectory,
 see Figure.\ref{fig:Lbeta}.
 
\begin{figure*}[tbp]h
 \begin{center}
  \begin{tabular}{cc}
 \resizebox{.49\textwidth}{!}
   {\includegraphics[scale=0.25,angle=270]{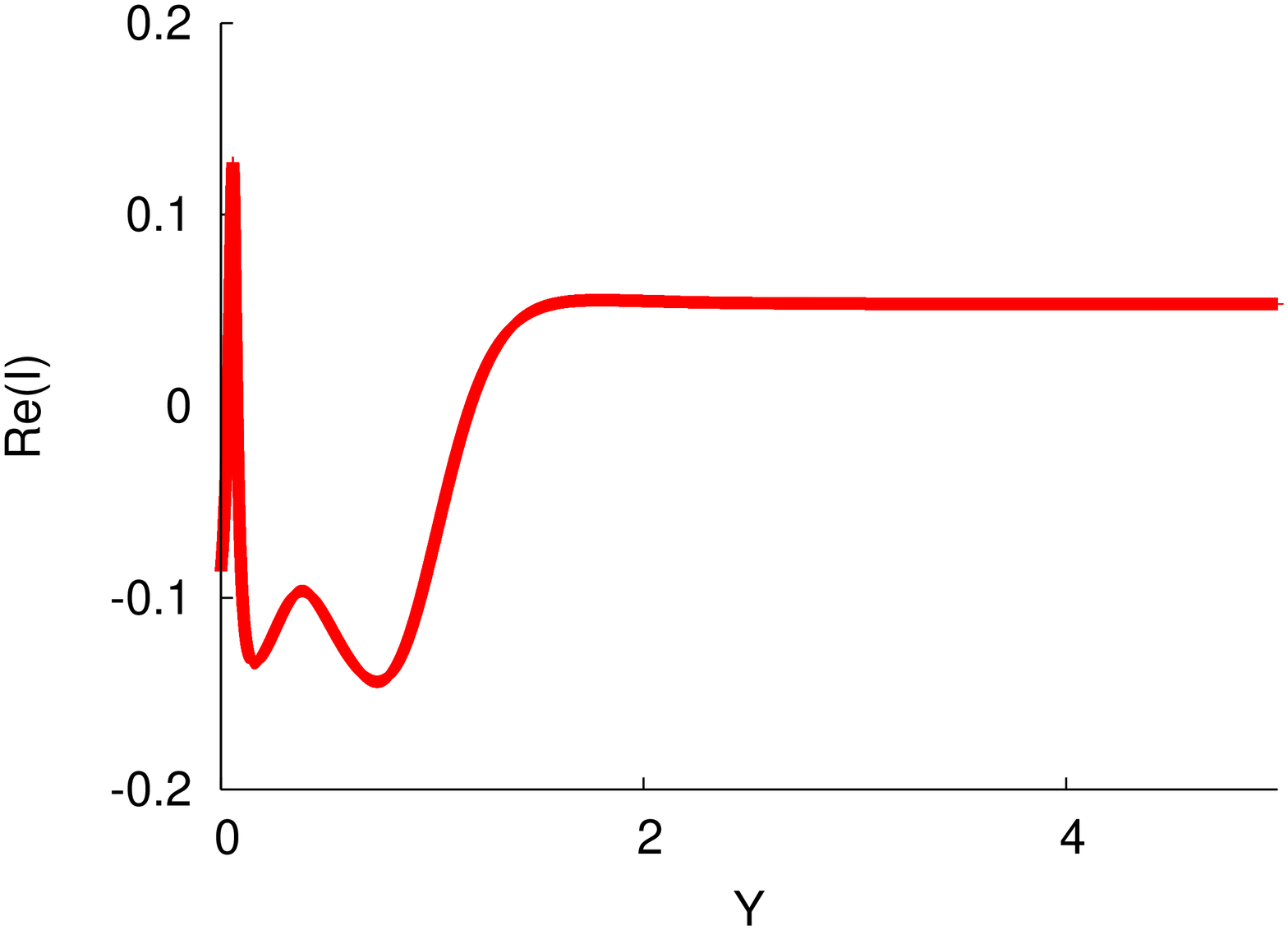}}& 
 \resizebox{.49\textwidth}{!}
   {\includegraphics[scale=0.25,angle=270]{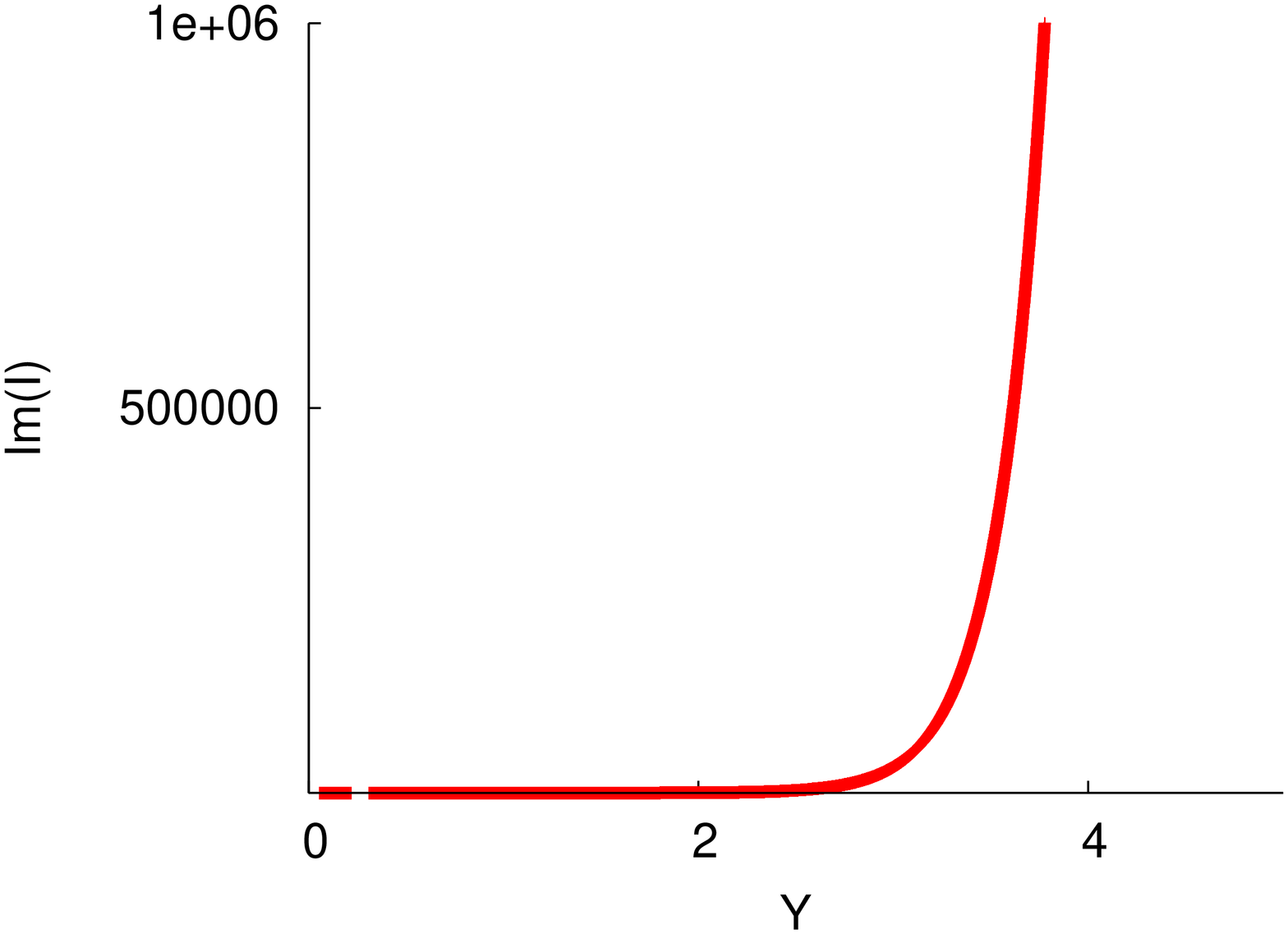}}
  \end{tabular}
  \caption{The real and imaginary part of
  the Euclidean action $I$.\newline
  \label{fig:Eaction}}
 \end{center}
\end{figure*}

The real and imaginary parts of the Euclidean action are shown in Figure.\ref{fig:Eaction}.
While the real part of the action is oscillating in $y<1$,  the real part stabilizes and imaginary part is rapidly 
varying in $y>2$ (in large scale).
This solution satisfies the classicality condition (\ref{CC}), thus
there exist a corresponding classical Lorentzian history.
The real part of Euclidean action provides the relative probability of
history.
\begin{equation}
  \rho(q^A) = e^{-2I_R(q^A)}\nabla_nS\nonumber
\end{equation}
In Figure.\ref{fig:Eaction}, the real part of action stabilizes
at $I_R \sim 0.5$. In the isotropic model, $I_R \sim -0.1$. 
We can see that the probability of the anisotropic universe is 
less than that of isotropic one.


\subsection{CLASSICAL LORENTZIAN SOLUTIONS}

Next, we calculate the Lorentzian histories by using the complex solution as
initial condition.

\begin{figure*}[p]
 \begin{center}
  \begin{tabular}{cc}
 \resizebox{.4\textwidth}{!}{\includegraphics[scale=0.15,angle=270]{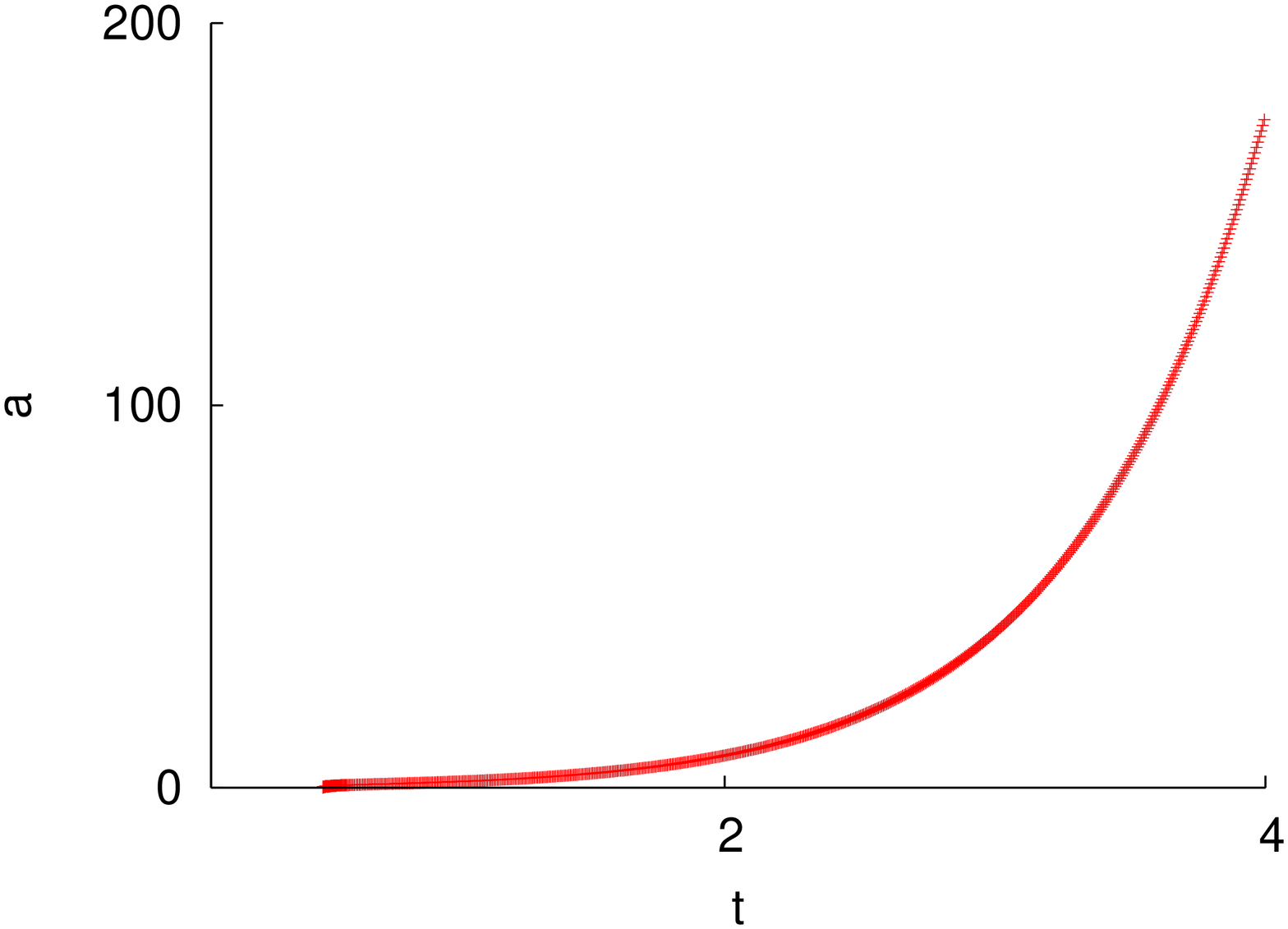}}& 
 \resizebox{.4\textwidth}{!}{\includegraphics[scale=0.15,angle=270]{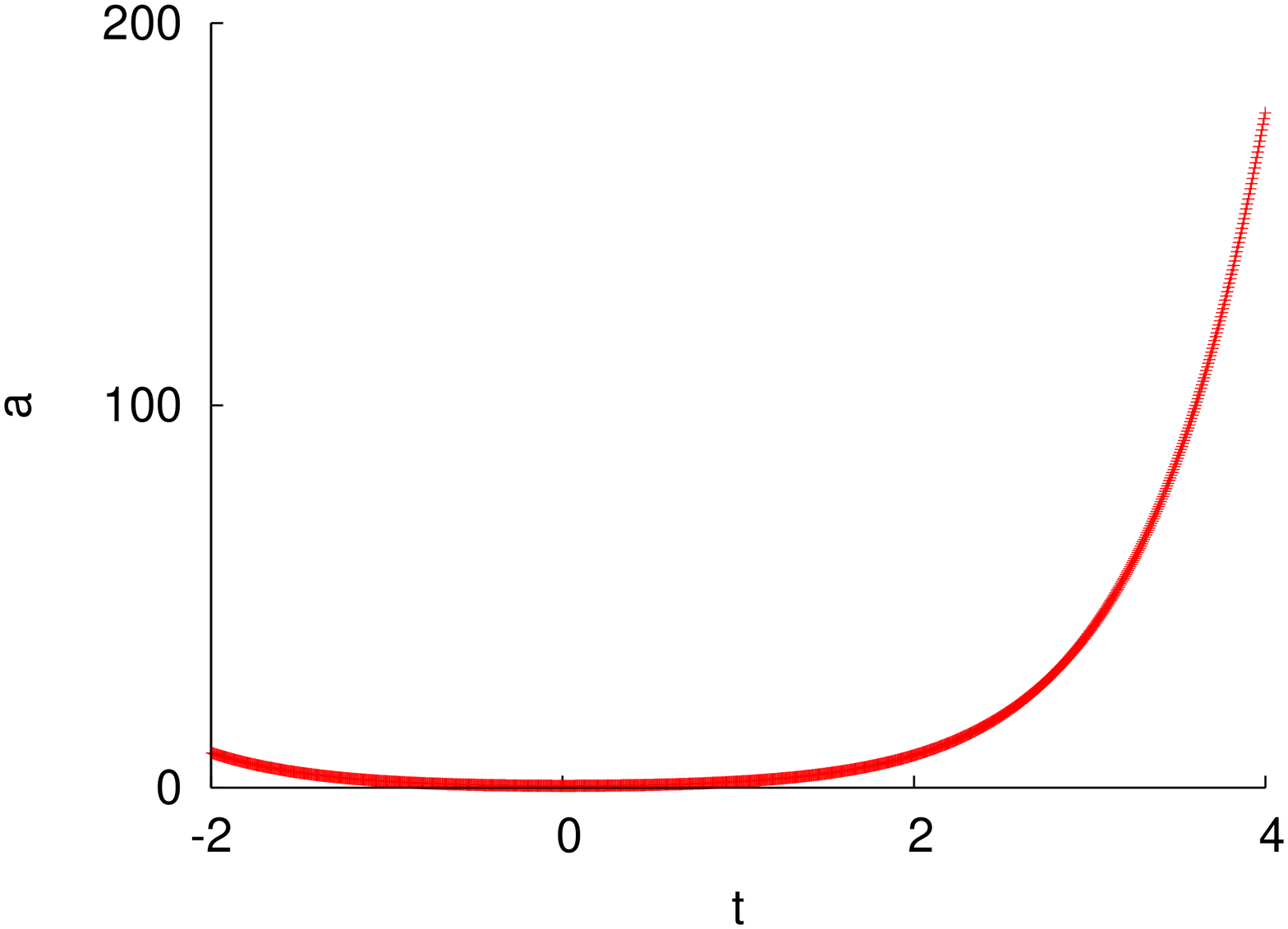}}\\
 \resizebox{.4\textwidth}{!}{\includegraphics[scale=0.15,angle=270]{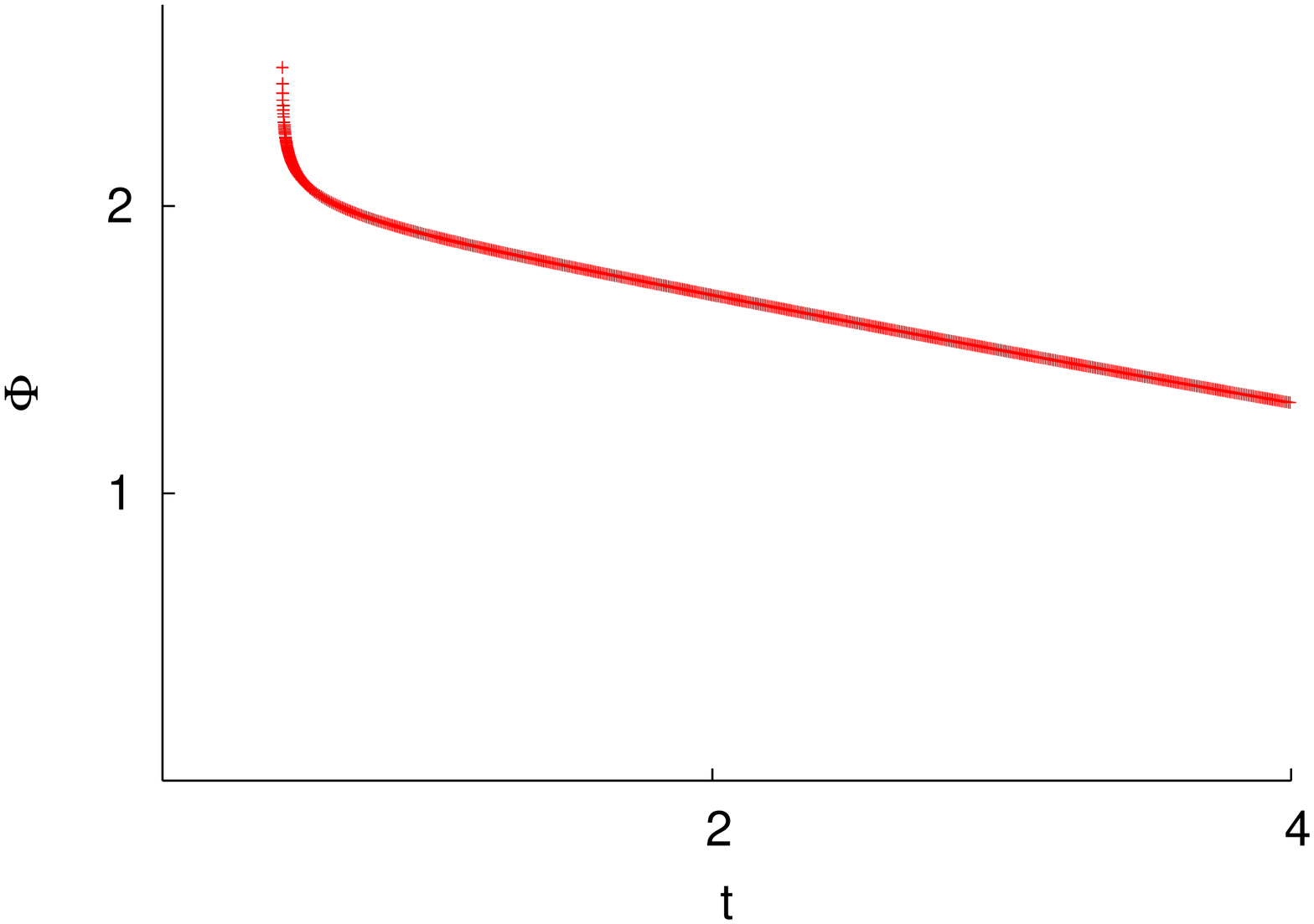}}&
 \resizebox{.4\textwidth}{!}{\includegraphics[scale=0.15,angle=270]{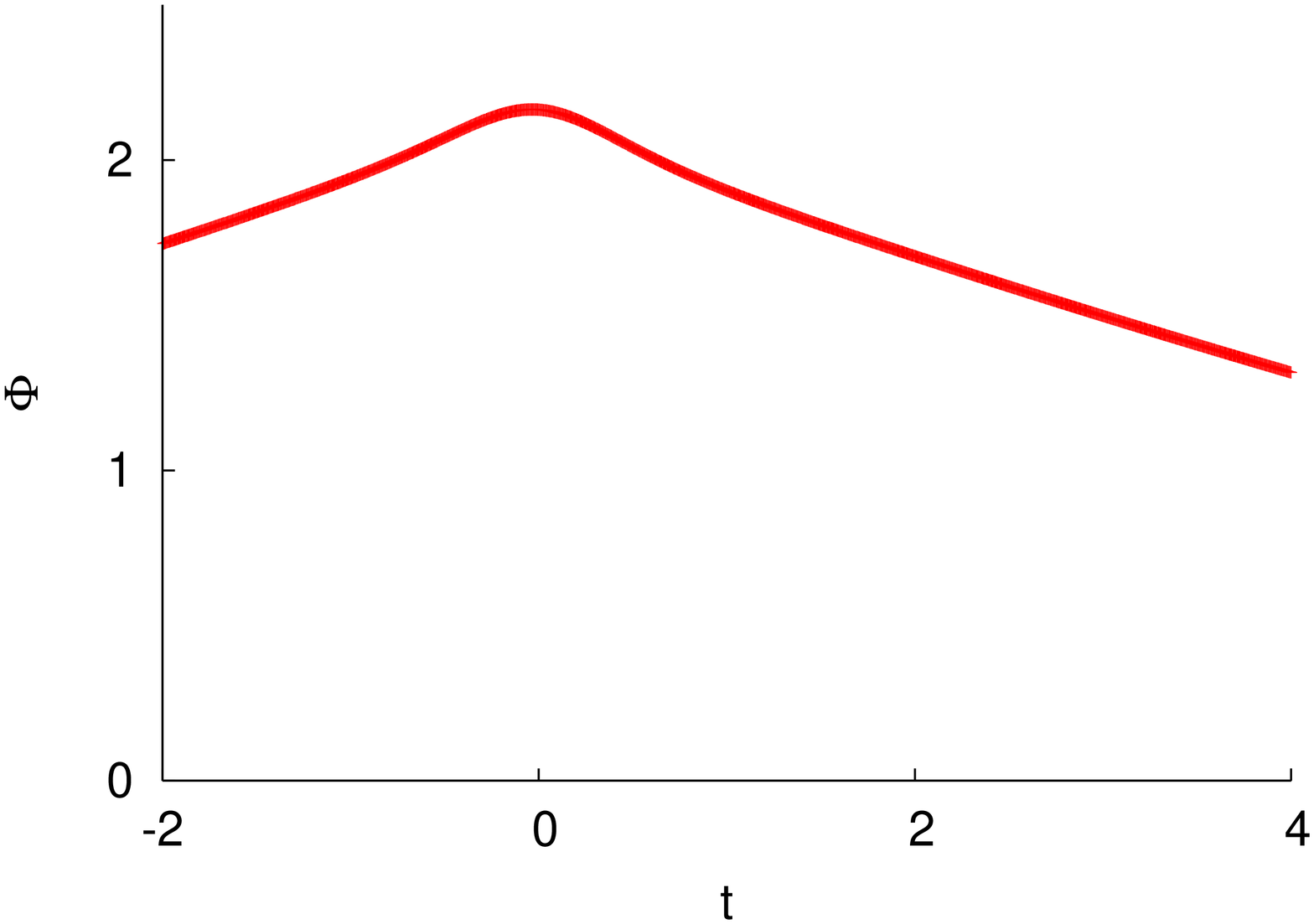}}\\
 \resizebox{.4\textwidth}{!}{\includegraphics[scale=0.15,angle=270]{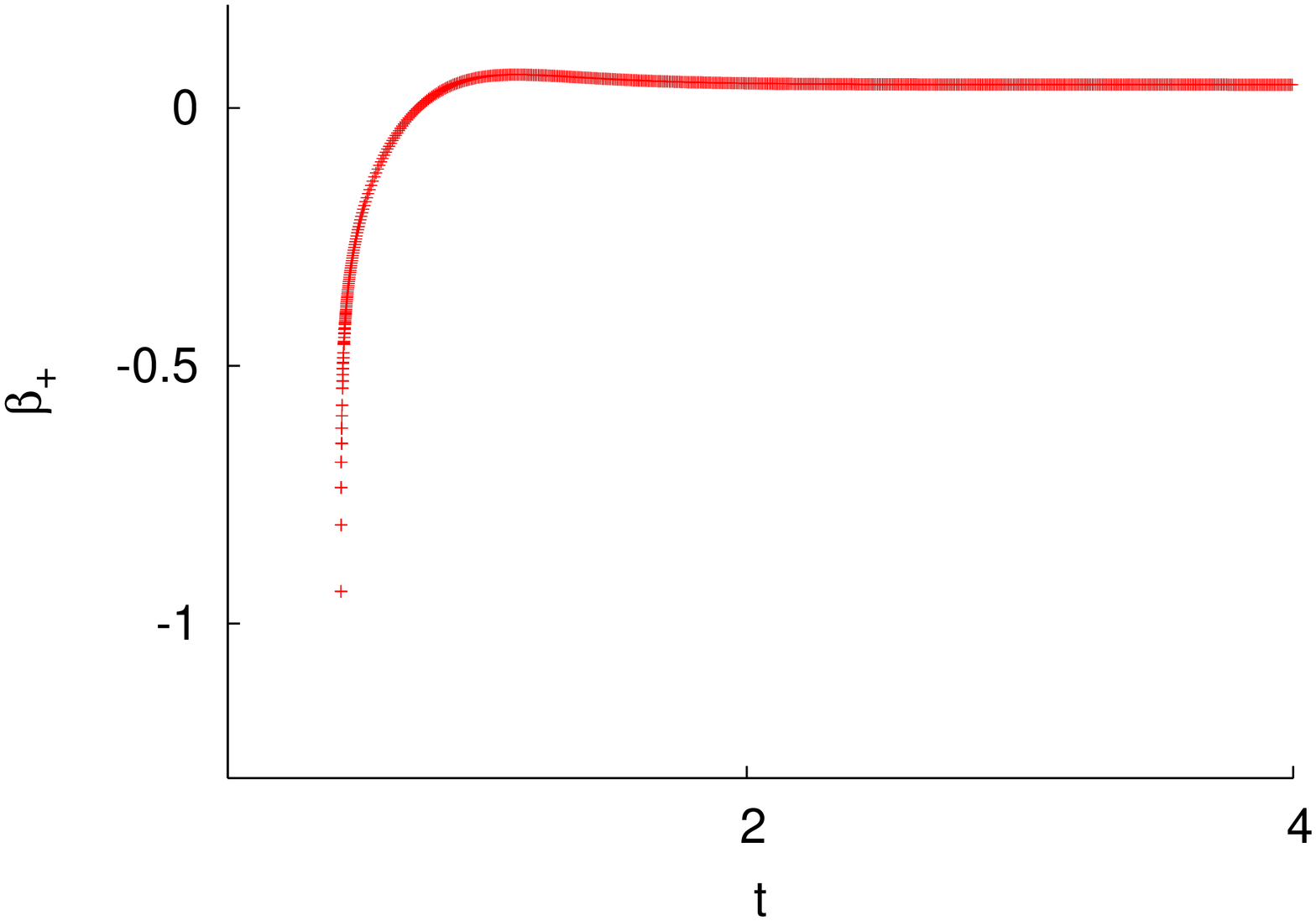}}& 
 \resizebox{.4\textwidth}{!}{\includegraphics[scale=0.15,angle=270]{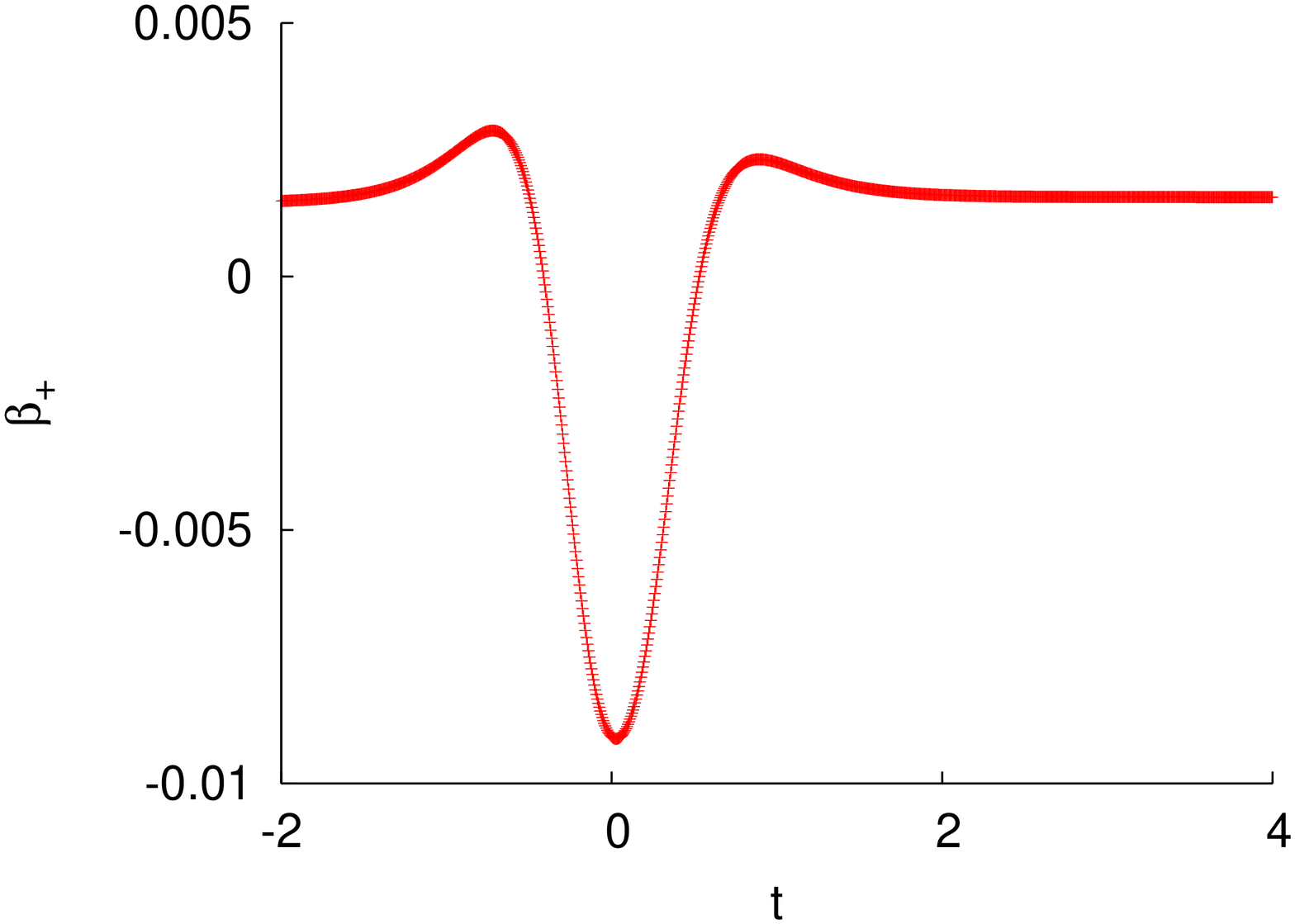}}\\
 \resizebox{.4\textwidth}{!}{\includegraphics[scale=0.15,angle=270]{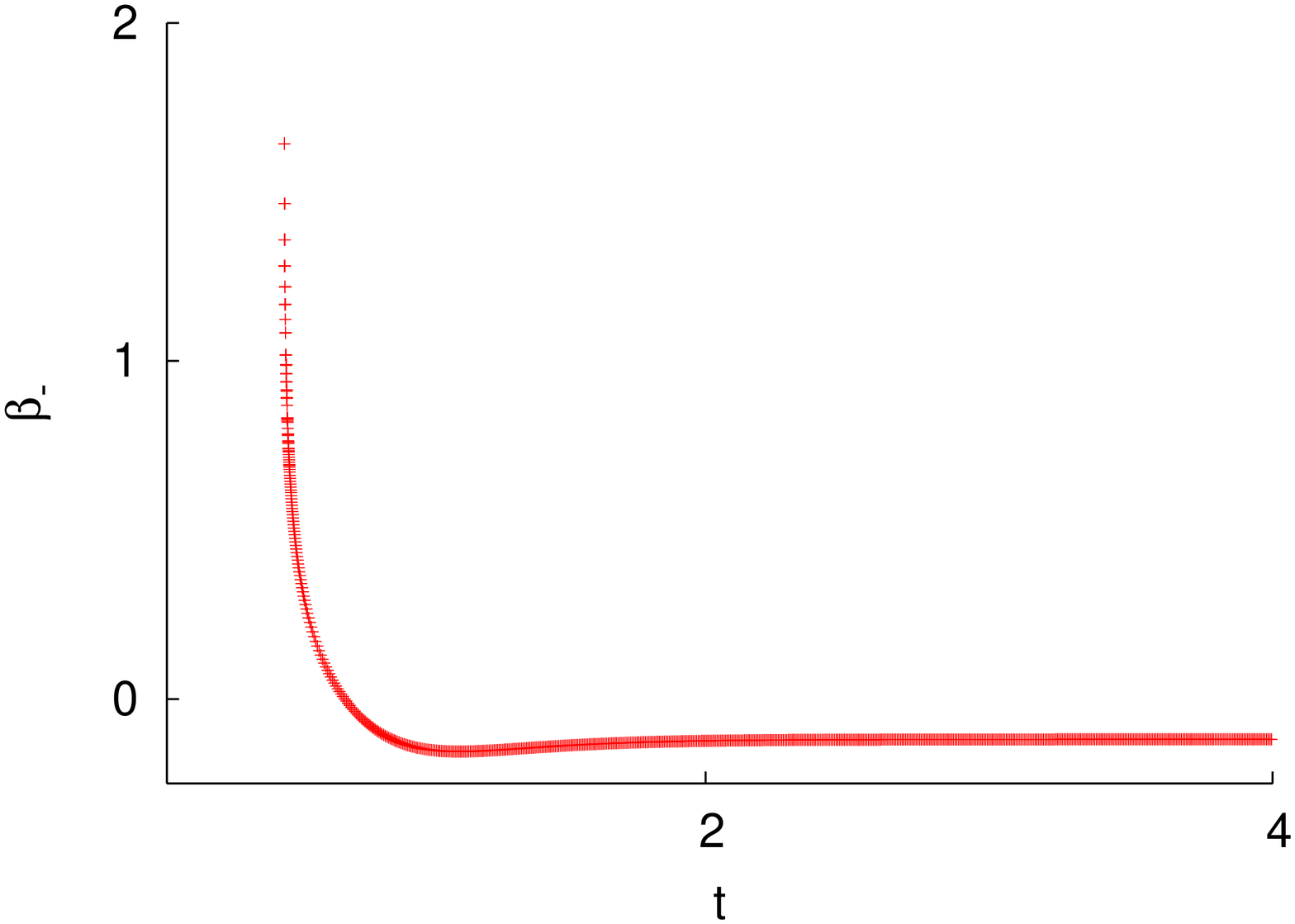}}&
 \resizebox{.4\textwidth}{!}{\includegraphics[scale=0.15,angle=270]{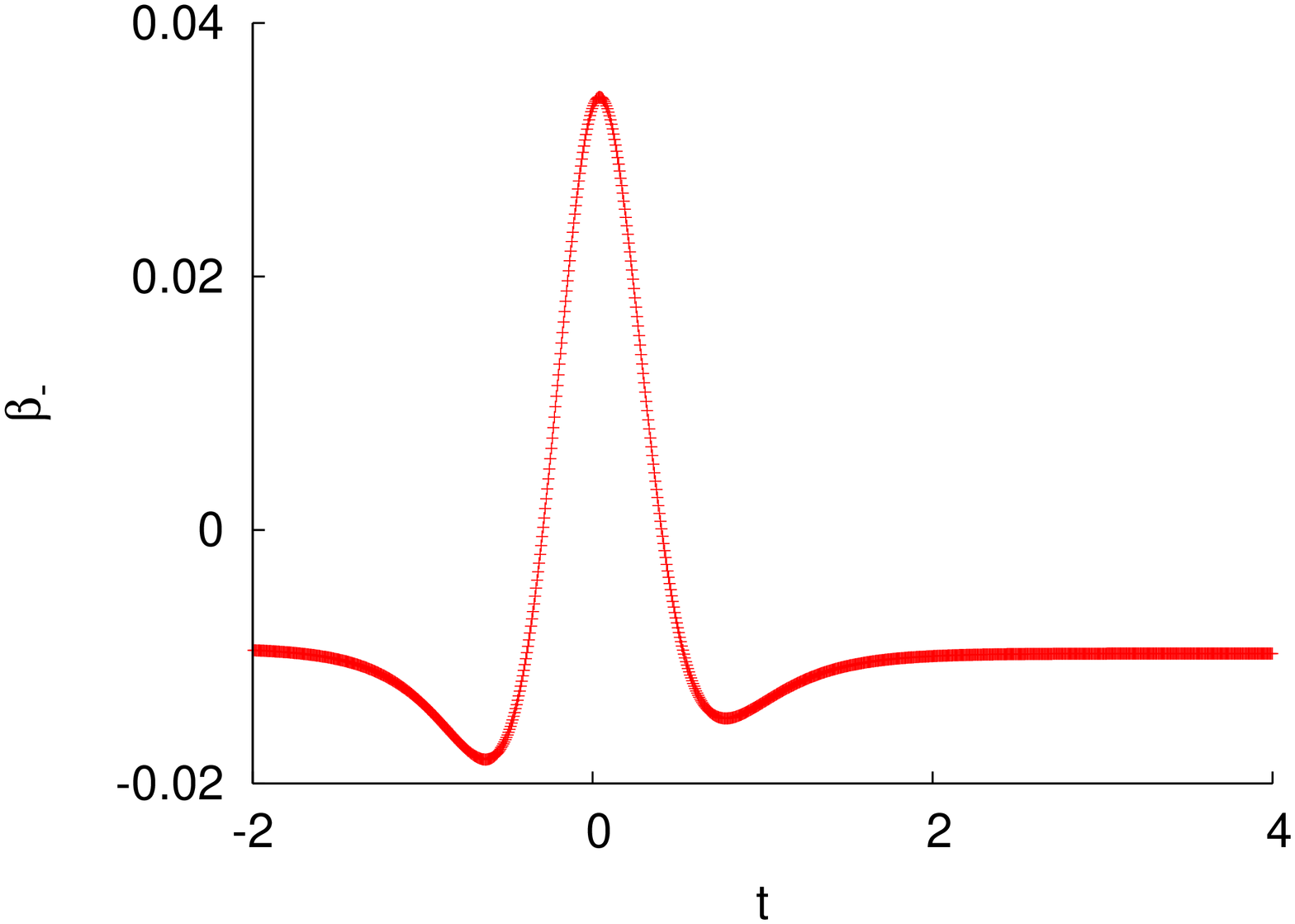}}\\
   \end{tabular}
  \caption{Lorentzian history of two models. 
  From top to bottom, $a,\phi,\beta_+,\beta_-$.\newline\hspace{3.4em}
  Left row is model A (high anisotropy), right row is model B (little
  anisotropy).\newline\hspace{3.4em}
  We use the complex Euclidean solution at $Y=4$ as initial value
  at $t=4$.\newline\hspace{3.4em}
  Model A: $\mu=3/4, \phi_0=2, P2=0.02, M2=0.15$ \newline\hspace{3.4em}
  Model B: $\mu=3/4, \phi_0=2, P2=0.002, M2=0.015$ 
  \label{fig:LX}}
\end{center}
\end{figure*}

Figure.\ref{fig:LX} shows the two Lorentzian histories. 
One (we call model A) has large anisotropy, the another (model B) has small anisotropy. 
If there is no anisotropy, the history shows initial bouncing behavior as shown by HHH. 
Similarly, model B is bouncing as in isotropic case. 
On the other hand, model A has an initial singularity. 
Usually, the existence of singularity is regarded as a sign of the 
breakdown of the theory. However HHH claimed that the no-boundary wave
function predicts the probability of classical  
{\it histories} rather than initial data,
thus the initial singularity is not a signal of break down of the theory.
Since the second time derivative of the scale factor $d^2a/dt^2$ is positive in
both models, these histories have inflationary period. 

The motion of anisotropy in triangular-shaped potential is shown in Figure.\ref{fig:Lbeta}.
Model A has started with high anisotropy, and then rolled down to narrow valley, and arrived
at low anisotropic state. Model B has started at the bottom area of potential, 
and wandering around the initial position.   
In both models, the anisotropy will approach to constant value at large
scale, and nearly vanishes. 
These features are consistent with preceding studies (e.g. Hawking and
Luttrell \cite{HandL}, Wright and Moss \cite{WandM} etc.).

\begin{figure*}[tbp]
 \begin{center}
  \begin{tabular}{cc}
 \resizebox{.49\textwidth}{!}
   {\includegraphics[scale=0.2,angle=270]{4a.eps}}& 
 \resizebox{.49\textwidth}{!}
   {\includegraphics[scale=0.2,angle=270]{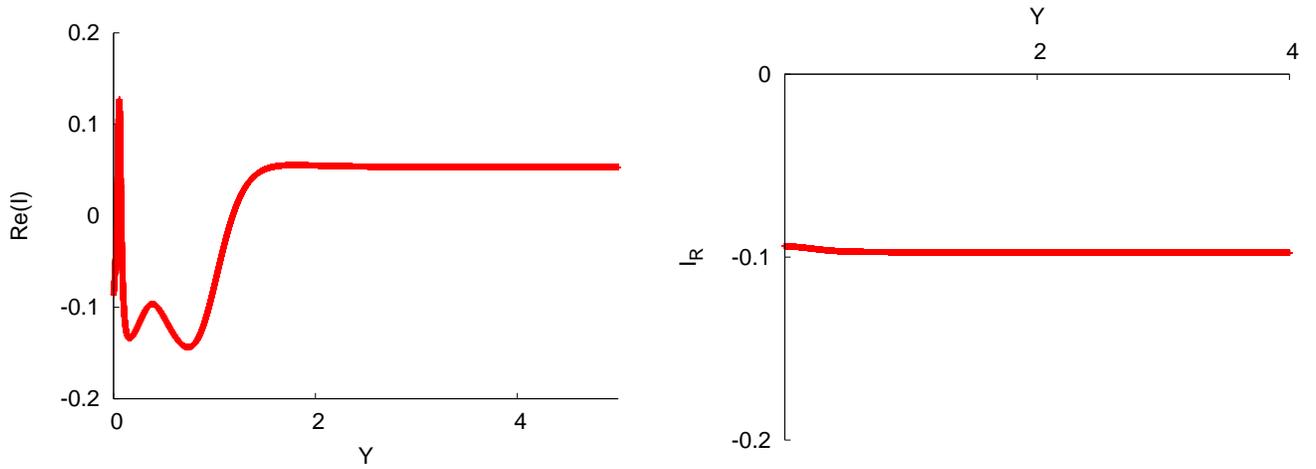}}
  \end{tabular}
  \caption{The real part of the Euclidean action.
  Left is model A, right is model B.\newline\hspace{3.4em}
  The probability of model A is smaller than model B.
  \label{fig:twoaction}}
 \end{center}
\end{figure*}

The real part of the Euclidean action is shown
in Figure.\ref{fig:twoaction}.
The figure clearly shows that the probability of high anisotropic
histories are smaller than low anisotropic histories. 
Model B's probability is slightly smaller
than isotropic one.\\

\begin{figure*}[tbp]
 \begin{center}
  \includegraphics[scale=0.5,angle=270]{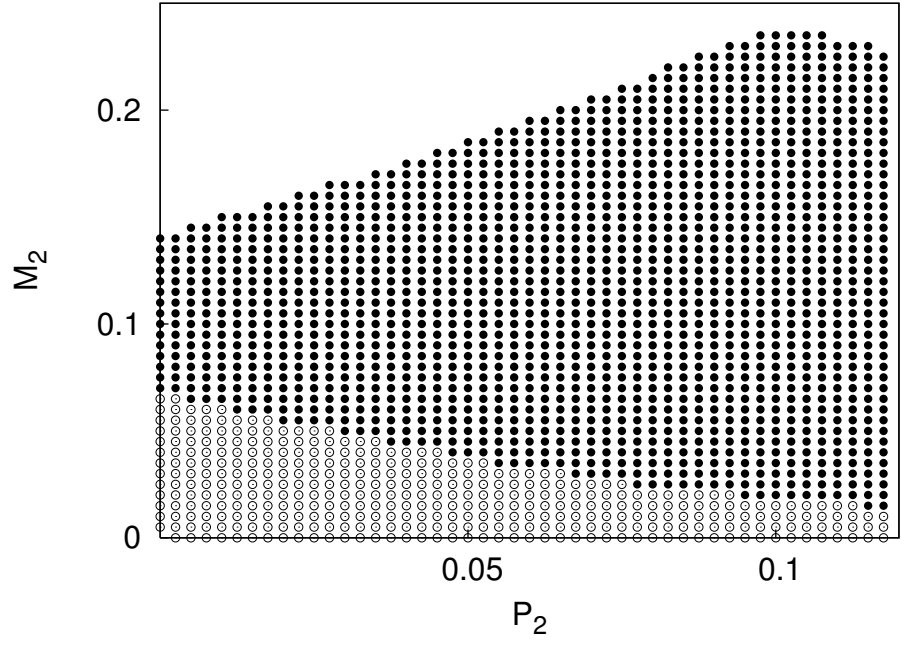}
  \end{center}
 \caption{The area where the classical histories exist 
 in $P_2 - M_2$ plane.\newline\hspace{3.4em}
 White points represents the histories which has initial bounce,
 black ones has initial singularity. \newline\hspace{3.4em}
 There is critical line 
 where the initial state changed and upper limits.
 \label{fig:area}}
\end{figure*}

\indent Figure.\ref{fig:area} shows the area 
where the classical histories exist in $P_2 - M_2$ plane.
We do not change $\mu,\phi_0$.
White points represent the histories with initial bounce, 
and black points represent histories with initial singularity.
Clearly we can see that there is the critical line where the 
initial state changed drastically, 
the high values of $P_2,M_2$ area have initial singularity.
Note that $P_2,M_2$ are the absolute values of 
{\it second differential} of $\beta_+,\beta_-$ in initial condition,
not the initial value. 
We could not find the solution outside of the plotted area. There seems to
be an upper limit in each $P_2,M_2$. 

\begin{figure*}[tbp]
 \begin{center}
  \begin{tabular}{cc}
 \resizebox{.49\textwidth}{!}
   {\includegraphics[scale=0.4,angle=270]{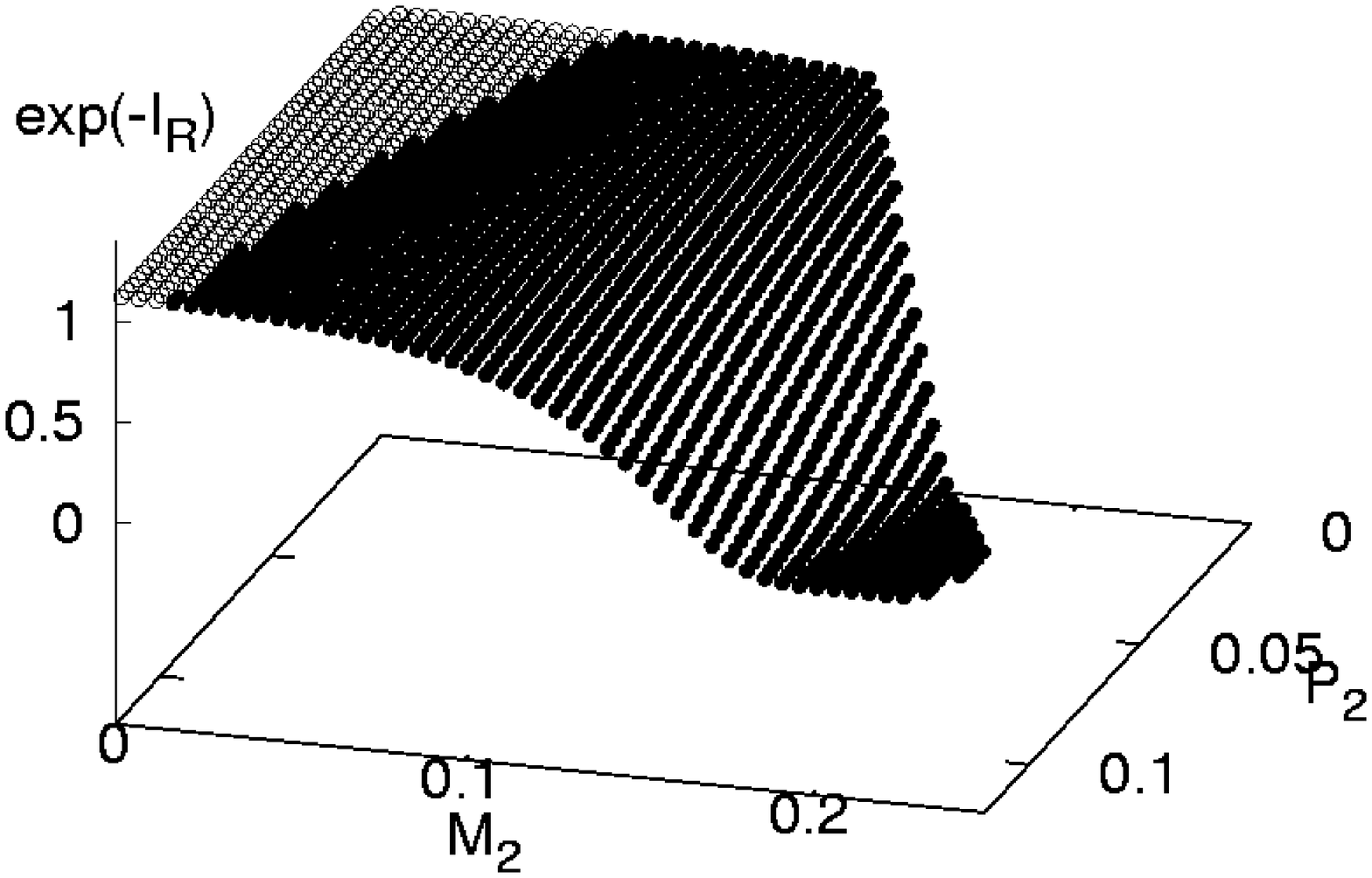}}& 
 \resizebox{.49\textwidth}{!}
   {\includegraphics[scale=0.4,angle=270]{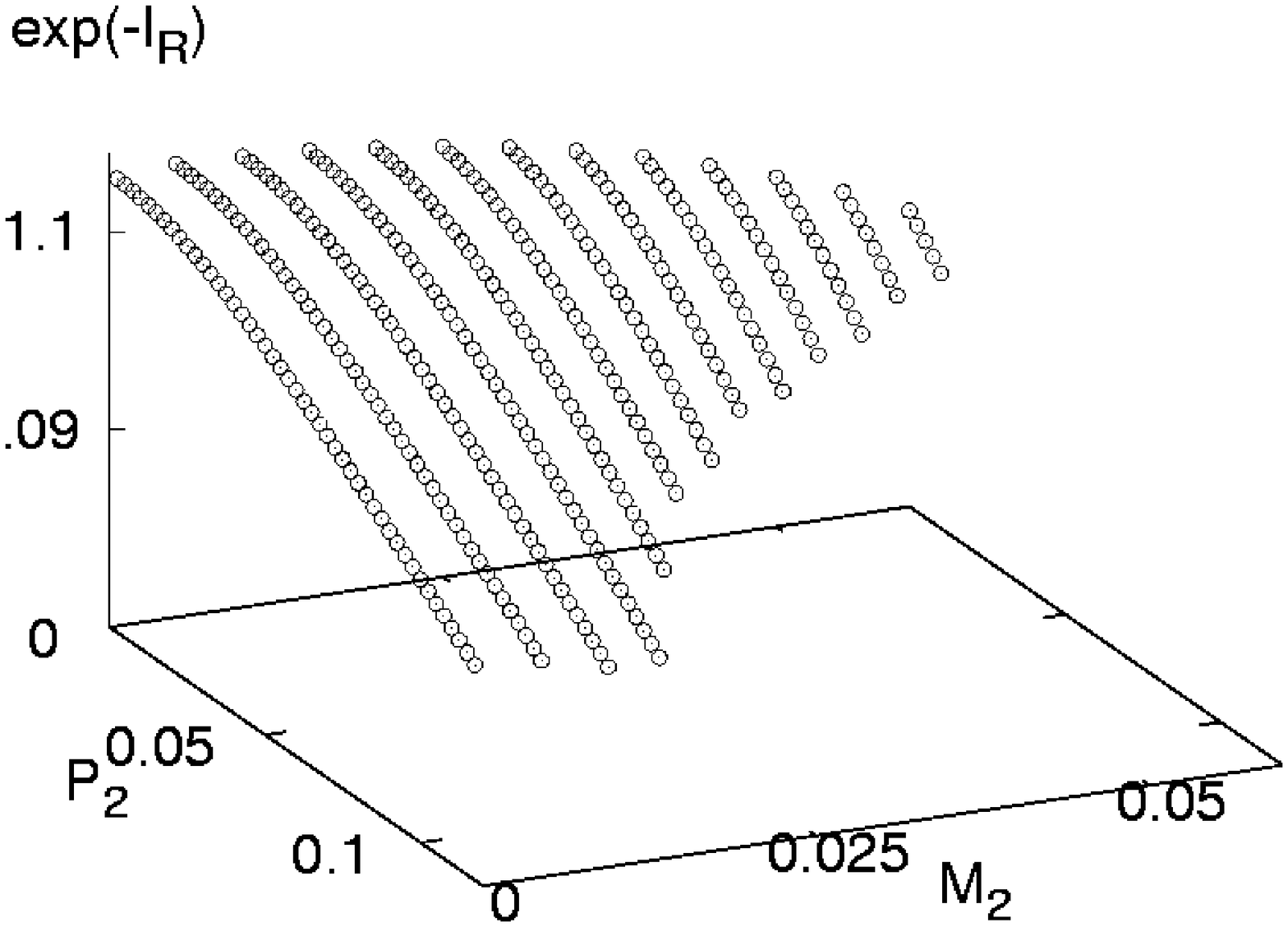}}
  \end{tabular}
  \caption{The relative probability distribution in $P_2 - M_2$ plane.
  \newline\hspace{3.4em}
  Left represents same area with Fig.\ref{fig:area}, right represents
  the area nearby origin (Isotropic). 
  \label{fig:distri}}
 \end{center}
\end{figure*}

\begin{figure*}[tbp]
 \begin{center}
  \includegraphics[scale=0.5,angle=270]{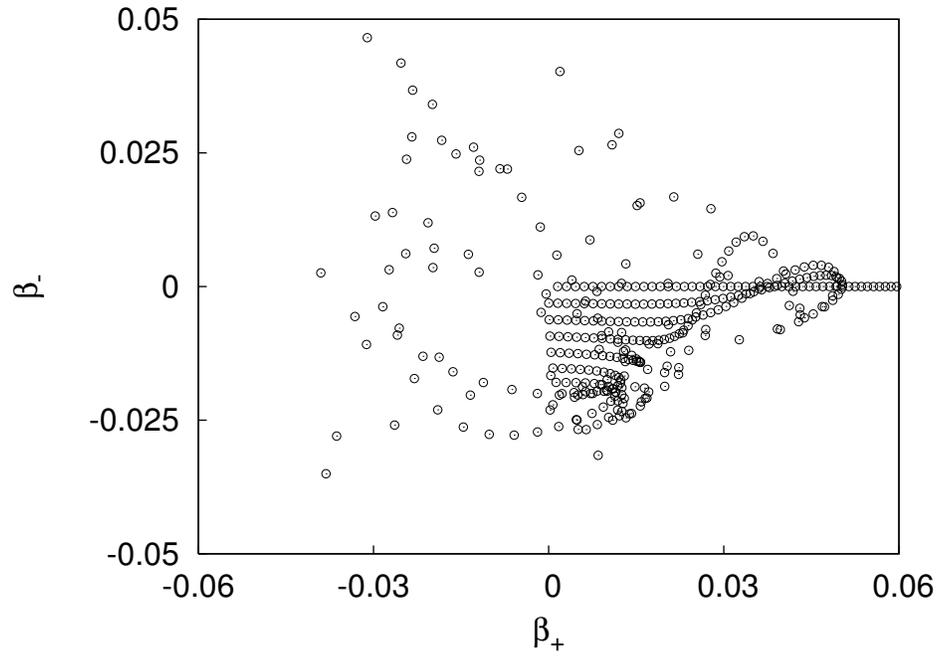}
  \end{center}
 \caption{$\beta_+$ vs $\beta_-$ at the bounce.
 \label{fig:Ldist}}
\end{figure*}

\begin{figure*}[tbp]
 \begin{center}
  \includegraphics[scale=0.5,angle=270]{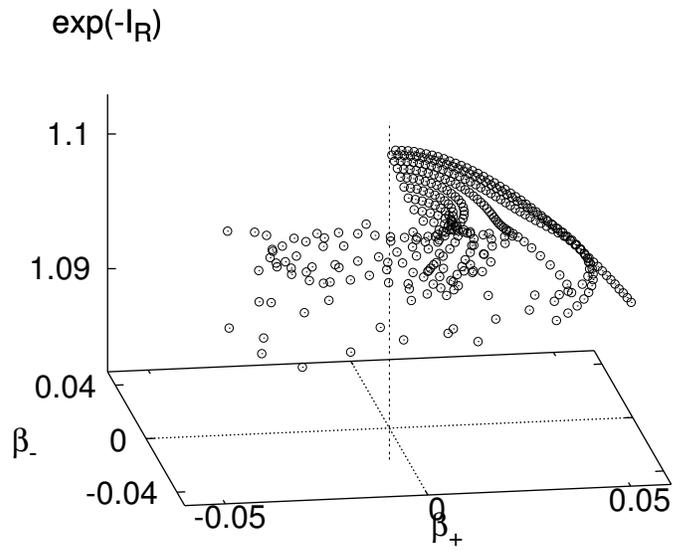}
  \end{center}
 \caption{The relative probability distribution in $\beta_+ - \beta_-$
 plane at the bounce.\newline\hspace{3.8em}
 There is high probability surface but no clear structure at low probability.
\label{fig:Lkakuritu}}
\end{figure*}

The relative probability distribution is 
shown in Figure.\ref{fig:distri}. 
The probability has maximum at isotropic history,
and is rapidly decreasing with the increase of anisotropy. 
Figure.\ref{fig:Ldist} shows the distribution of $\beta_+, \beta_-$
at the bounce, and Figure.\ref{fig:Lkakuritu} shows their probability distribution. 
We can see that states with low anisotropies have higher probabilities. 

We found that the distribution of the anisotropy near the bounce is highly
inhomogeneous. This is very different from classical result. 
While the classical result predicts chaotic behavior near the initial
singularity \cite{Misner}\cite{BKL}, we could not see such a behavior.
  
\begin{figure*}[tbp]
 \begin{center}
  \includegraphics[scale=0.5,angle=270]{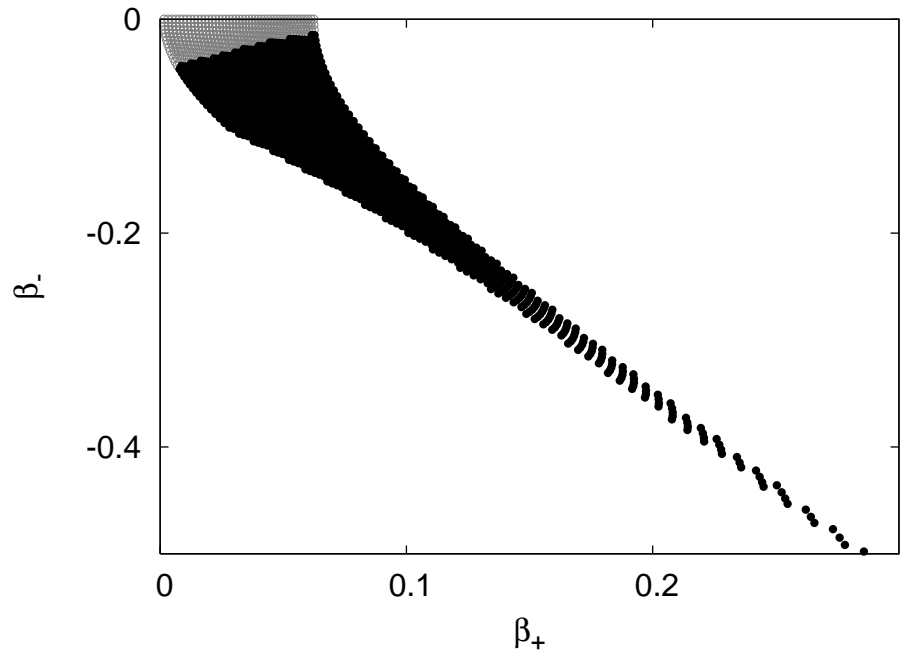}
  \end{center}
 \caption{$\beta_+$ vs $\beta_-$ at large scale in classical
 history.\newline
 \hspace{3.7em}
 White region represents bouncing history, black region has initial singularity.
 \newline
 \hspace{4em}Since the anisotropy is constant 
 at large scale (see Fig.~\ref{fig:LX}),
 this figure represents the observable one.
 \label{fig:bunpu}}
\end{figure*}

\begin{figure*}[tbp]
 \begin{center}
  \includegraphics[scale=0.5,angle=270]{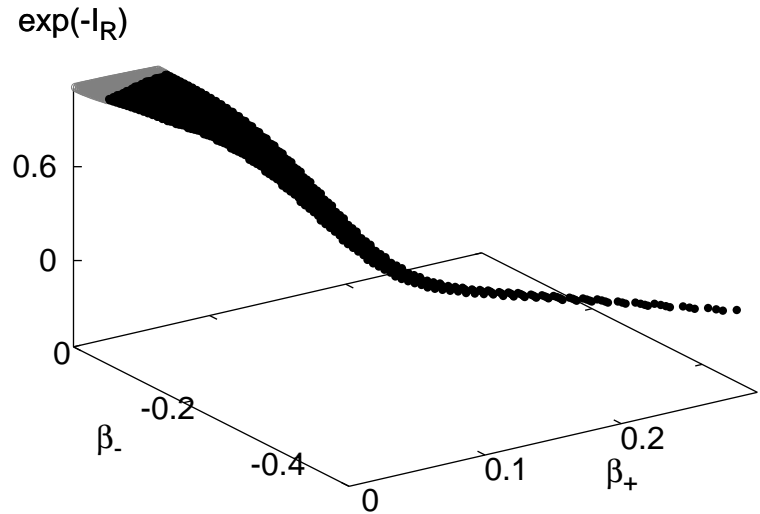}
  \end{center}
 \caption{The relative probability distribution in $\beta_+ - \beta_-$
 at large scale in classical history.\newline\hspace{3.82em}
 The probability distribution is peaked around the isotropic universe. 
\label{fig:Kbunpu}}
\end{figure*}

The anisotropy at large scale are shown 
in Figure.\ref{fig:bunpu} and Figure.\ref{fig:Kbunpu}. 
Clearly, the probability is peaked around the isotropic history.
Therefore {\it the no-boundary wave function of the universe
predicts nearly isotropic universe.} 

\section{CONCLUSION \& DISCUSSION}

We investigated the emergence of 
the classical Mixmaster universe with a single scalar field
and with the cosmological constant from the no-boundary quantum state according to the method of HHH. 
We confirmed that there are complex solutions that 
satisfy the classicality condition, thus there are
classical Lorentzian histories with any amount of anisotropies.
This fact implies that the method of HHH can be applied for a wide range of models.
It is also found that the probability of anisotropic universe is lower than isotropic one. 
In addition the emerged classical universe experiences the inflationary expansion and the remaining anisotropy decreases rapidly. 
Thus the quantum cosmology with no-boundary wave function may be able to explain quite naturally the emergence of our universe. 

We also found that the classical histories have initial singularity or initial
bounce depending on the initial acceleration of anisotropies. High acceleration may induce the initial singularity.
The no-boundary proposal seems to predict that the distribution of anisotropy near the bounce or singularity 
is not homogeneous and very different from the prediction by classical
theory and by analysis of Wheeler-DeWitt equation \cite{MRG}.  
On the other hand, the loop quantum cosmology predicts non-chaotic 
behavior near singularity \cite{BDH}. 
Thus it would be very interesting to see if the non-chaotic behavior is general 
in more general spacetimes. 

\begin{acknowledgments}
We thank S.~Yoshida for useful advices of numerical calculation. 
This work is supported in part by a Grants-in-Aid for 
Scientific Research from JSPA(Nos. 1807200, 20540245) 
as well as by Core-to-Core 
Program "International Research Network for Dark Energy".
\end{acknowledgments}

\newpage

\end{document}